\documentclass{article}

\usepackage{arxiv}

\usepackage[utf8]{inputenc} 
\usepackage[T1]{fontenc}    
\usepackage{hyperref}       
\usepackage{url}            
\usepackage{booktabs}       
\usepackage{amssymb}
\usepackage{amsmath}
\usepackage{nicefrac}       
\usepackage{microtype}      
\usepackage{lipsum}		
\usepackage{pdflscape}
\usepackage{booktabs}
\usepackage{graphicx}
\usepackage{subcaption}
\usepackage{longtable}
\usepackage{tabularx}
\usepackage{enumitem}
\usepackage[table]{xcolor}
\usepackage{multirow} 
\usepackage{array}
\usepackage{url}
\usepackage[numbers]{natbib}
\usepackage{doi}

\title{Decision Support System for Technology Opportunity Discovery: An Application of the Schwartz Theory of Basic Values}

\author{ \href{https://orcid.org/0000-0002-4774-1506}{\includegraphics[scale=0.06]{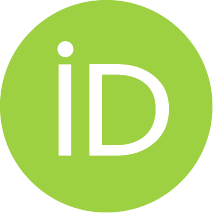}\hspace{1mm}Ayato Kitadai}\thanks{\texttt{a.kitadai@css.t.u-tokyo.ac.jp}}\\
    School of Engineering\\
	The University of Tokyo\\\\
	\And
	Takumi Ito \\
    School of Engineering\\
	The University of Tokyo\\
    \And
	Yumiko Nagoh \\
    School of Engineering\\
	The University of Tokyo\\
	\And
	\href{https://orcid.org/0000-0003-1356-1467}{\includegraphics[scale=0.06]{orcid.pdf}\hspace{1mm}Hiroki Takahashi} \\
    Faculty of Engineering\\
	University of Tsukuba\\
    \And
	\href{https://orcid.org/0000-0003-1543-3303}{\includegraphics[scale=0.06]{orcid.pdf}\hspace{1mm}Masanori Fujita} \\
    College of International Management\\
	Ritsumeikan Asia Pacific University\\
    \And
	\href{https://orcid.org/0000-0003-2864-5874}{\includegraphics[scale=0.06]{orcid.pdf}\hspace{1mm}Sangjic Lee}\\
    College of Commerce\\
	Nihon University\\
    \And
	Fumiaki Miyahara \\
    Sony Computer Science Laboratories, Inc.\\
    \And
	Tetsu Natsume \\
    Sony Computer Science Laboratories, Inc.\\
    \And
	\href{https://orcid.org/0000-0002-6411-8716}{\includegraphics[scale=0.06]{orcid.pdf}\hspace{1mm}Nariaki Nishino} \\
    School of Engineering\\
	The University of Tokyo\\
}

\renewcommand{\shorttitle}{DSS for TOD: An Application of the Schwartz Theory of Basic Values}

\hypersetup{
pdftitle={Decision Support System for Technology Opportunity Discovery: An Application of the Schwartz Theory of Basic Values},
pdfsubject={q-bio.NC, q-bio.QM},
pdfauthor={Ayato Kitadai},
pdfkeywords={Technology Opportunity Discovery, Schwartz Theory of Basic Values, Technology Readiness Level, R\&D},
}

\begin{document}
\maketitle

\begin{abstract}
Discovering technology opportunities (TOD) remains a critical challenge for innovation management, especially in early-stage development where consumer needs are often unclear.
Existing methods frequently fail to systematically incorporate end-user perspectives, resulting in a misalignment between technological potentials and market relevance.
This study proposes a novel decision support framework that bridges this gap by linking technological feasibility with fundamental human values.
The framework integrates two distinct lenses: the engineering-based Technology Readiness Levels (TRL) and Schwartz's theory of basic human values.
By combining these, the approach enables a structured exploration of how emerging technologies may satisfy diverse user motivations.
To illustrate the framework's feasibility and insight potential, we conducted exploratory workshops with general consumers and internal experts at Sony Computer Science Laboratories, Inc., analyzing four real-world technologies (two commercial successes and two failures).
Two consistent patterns emerged: (1) internal experts identified a wider value landscape than consumers (``\textit{vision gap}''), and (2) successful technologies exhibited a broader range of associated human values (``\textit{value breadth}''), suggesting strategic foresight may underpin market success.
This study contributes both a practical tool for early-stage R\&D decision-making and a theoretical link between value theory and innovation outcomes.
While exploratory in scope, the findings highlight the promise of value-centric evaluation as a foundation for more human-centered technology opportunity discovery.
\end{abstract}

\keywords{Technology Opportunity Discovery \and Schwartz Theory of Basic Values \and Technology Readiness Level \and R\&D}

\section{Introduction}
Technology opportunity discovery (TOD) is a critical task for organizations seeking to identify promising avenues for technological development.
Defined as the potential for technological progress~\cite{olsson_2005}, TOD significantly shapes firms' R\&D strategies and long-term competitiveness~\cite{klevorick_1995}.
However, despite decades of research, effectively aligning emerging technologies with real market needs remains a persistent challenge~\cite{markman_2009}.

Existing approaches to TOD fall into three broad categories: technology-based, firm-based, and market-based~\cite{yun_2021}.
Technology-based methods rely on technical data such as patents~\cite{wang_2019, wang_2022, yun_2022, yoon_2015, park_2017, park_2022, lee_2022, li_2023, choi_2023, zhao_2025} to assess novelty; firm-based methods analyze organizational capabilities~\cite{yoon_2017}; and market-based methods, such as technology roadmapping or new product development frameworks~\cite{phaal_2004, kerr_2021, zhang_2019, jin_2015}, attempt to integrate user perspectives.
Yet these methods often rest on predefined application assumptions, which may constrain exploration and overlook latent consumer needs.
This methodological gap often leads to poor forecasts of a technology's social acceptance and eventual market relevance.

In recent years, scholars have emphasized the importance of incorporating end-user values into the early stages of technological planning~\cite{carr_1992, lo_2012}.
While it is widely recognized that consumers' perceived value plays a pivotal role in commercialization success, few methodologies systematically map these perceived values to technological capabilities.
In particular, the alignment between emerging technologies and fundamental human motivations has received limited attention in TOD literature~\cite{spitsberg_2013}.

This study addresses this gap by introducing a novel decision support framework for early-stage TOD.
Our approach integrates two distinct perspectives: technological feasibility, as assessed through Technology Readiness Levels (TRL), and user-centric value perception, grounded in Schwartz's theory of basic human values~\cite{schwartz_1992}.
By explicitly mapping technology functions to user-perceived values, the framework facilitates structured, human-centric exploration of potential use cases for emerging technologies.

To illustrate the framework's feasibility, we conducted exploratory workshops with general consumers and internal experts at Sony Computer Science Laboratories, Inc. (hereafter Sony CSL), examining four real-world technologies (two successful and two unsuccessful).
While the scope of this study is intentionally limited, it highlights two recurring patterns—what we term the ``\textit{vision gap}'' and ``\textit{value breadth}''—that may serve as early indicators of commercialization potential.
 
This paper contributes to both theory and practice in decision support.
Theoretically, it introduces a novel integration of Schwartz's value theory into the technology opportunity discovery process, offering a structured method to assess the alignment between technological functions and fundamental human motivations.
This operationalization extends the application of psychological value theory into the domain of innovation management and early-stage R\&D strategy.

Practically, the study presents a workshop-based framework that enables decision-makers such as R\&D planners and innovation managers to systematically explore latent user needs and evaluate opportunity potential from both market and technical perspectives.
By revealing two actionable indicators--``\textit{vision gap}'' and ``\textit{value breadth}''--the framework supports more informed, value-driven decisions about which technologies to advance toward commercialization.

The remainder of this paper is structured as follows. Section~\ref{sec:theoretical_background} reviews the TRL concept and Schwartz's value theory. Section~\ref{sec:method} presents the proposed methodology.
Section~\ref{sec:workshop} describes the workshop-based evaluation, followed by the results and discussion in Section~\ref{sec:result} and \ref{sec:discussion}.
Section~\ref{sec:conclusion} concludes with implications, limitations, and directions for future research.

\section{Theoretical Background}\label{sec:theoretical_background}

To systematically discover technology opportunities, we argue that three distinct but interconnected concepts must be considered. 
First, we discuss the Technology Readiness Level (TRL) as a scale for objectively assessing the maturity and feasibility of a technology.
Next, we explore the concept of function modeling in engineering design, in which the function serves as a crucial bridge linking a technology's objective capabilities to subjective human experiences.
Finally, we introduce the Schwartz theory of basic values as a comprehensive framework for understanding the fundamental human needs that functions must ultimately satisfy.
We demonstrate how integrating these three concepts provides a scientifically rigorous foundation for our technology opportunity discovery framework.

\subsection{Technology Readiness Level: Measuring Technological Maturity}

A critical first step in technology opportunity discovery is to objectively evaluate feasibility and maturity of technologies.
The TRL is a globally recognized metric for this purpose, providing a systematic scale to assess a technology's progress from basic research to operational deployment.

The concept of technological maturity has a rich history, as reviewed by Mihaly~\cite{mihaly_2017}. Its development dates back to 1969, when NASA, planning for the post-Apollo era, required a method to review the readiness of technologies for future space missions.
This idea was formally documented in 1989 as ``readiness levels'' in a paper by NASA Headquarters~\cite{sadin_1989}.
At a time when technology was developed in an opportunity-driven manner, this scale was essential for assessing the feasibility of new technologies for actual space missions.
This led to the establishment of an initial seven-level scale (Table~\ref{tab:NASA7_TRLs}).

\begin{table}[ht]
    \centering
    \caption{Initial 7 Readiness Levels Described by NASA}
    \label{tab:NASA7_TRLs}
    \begin{tabular}{@{}ll@{}}
        \toprule 
        \textbf{Level} & \textbf{Description} \\
        \midrule 
        1 & Basic principles observed and reported \\
        2 & Potential application validated \\
        3 & Proof-of-Concept demonstrated analytically and/or experimentally \\
        4 & Component and/or breadboard laboratory validated \\
        5 & Component and/or breadboard validated in simulated or real-space environment \\
        6 & System adequacy validated in simulated environment \\
        7 & System adequacy validated in space \\
        \bottomrule 
    \end{tabular}
\end{table}

This scoring proved highly beneficial to NASA, enabling them to minimize mission risks and build a basis for future planning.
The scale was later expanded to the more generalized nine levels, for which Mankins~\cite{mankins_1995} provided a descriptive discussion.
It is now widely used not only by NASA but also by other institutions like the European Space Agency (ESA) and the US Department of Defense (DoD) (Table~\ref{tab:NASA9_TRLs}).

\begin{table}[ht]
    \centering
    \caption{Nine technology readiness levels (NASA definition)}
    \label{tab:NASA9_TRLs}
    \begin{tabularx}{\textwidth}{@{}lX@{}}
        \toprule
        \textbf{Level} & \textbf{Description} \\
        \midrule 
        1 & Basic principles observed and reported \\
        2 & Technology concept and/or application formulated \\
        3 & Analytical and experimental critical function and/or characteristic proof-of-concept \\
        4 & Component/subsystem validation in a laboratory environment \\
        5 & System/subsystem/component validation in a relevant environment \\
        6 & System/subsystem model or prototyping demonstration in a relevant end-to-end environment (ground or space) \\
        7 & System prototyping demonstration in an operational environment (ground or space) \\
        8 & Actual system completed and ``mission qualified'' through test and demonstration in an operational environment (ground or space) \\
        9 & Actual system ``mission proven'' through successful mission operations (ground or space) \\
        \bottomrule
    \end{tabularx}
\end{table}

Furthermore, as the EU promoted Key Enabling Technologies (KETs) to ensure economic competitiveness, the TRL framework was adapted to assess the market readiness of industrial technologies, notably in the Horizon2020 program (Table~\ref{tab:EU_TRLs}).

\begin{table}[ht]
    \centering
    \caption{Readiness Levels Used in the EU Horizon 2020}
    \label{tab:EU_TRLs}
    \begin{tabularx}{\textwidth}{@{}lX@{}}
        \toprule
        \textbf{Level} & \textbf{Description} \\
        \midrule
        1 & Basic principles observed \\
        2 & Technology concept formulated \\
        3 & Experimental proof of concept \\
        4 & Technology validated in laboratory \\
        5 & Technology validated in relevant environment (industrially relevant environment in the case of key enabling technologies) \\
        6 & Technology demonstrated in relevant environment (industrially relevant environment in the case of key enabling technologies) \\
        7 & System prototype demonstration in operational environment \\
        8 & System complete and qualified \\
        9 & Actual system proven in operational environment (competitive manufacturing in the case of key enabling technologies or in space) \\
        \bottomrule
    \end{tabularx}
\end{table}

The utility of TRL, however, extends beyond a simple maturity assessment.
Nakamura et al.~\cite{nakamura_2012}, for instance, combined TRL with the multi-level perspectives (MLP) framework to analyze innovation in the aviation industry.
Their work suggests that integrating TRL with other, non-technical factors can enhance our understanding of how technology succeeds or fails in a socio-technical context.

Building on this idea, our study utilizes TRL not just as a standalone measure of maturity, but as a crucial axis for evaluating the feasibility of achieving the specific values perceived by consumers.
It is for this purpose---linking technological readiness to value realization---that we developed our TRL evaluation scale, adapting the well-established EU Horizon 2020 framework.

\subsection{Functions: Bridging Technology and Human Intention}\label{subsec:functions}

A technology itself does not directly create value for consumers; rather, its value is realized when the functions that the technology provides are used by consumers in a specific environment.
Therefore, to analyze the link between a technology and consumer value, it is necessary to deconstruct the technology into a set of distinct functions.

The concept of ``function'' has been a central topic in engineering design.
A seminal work is the Function-Behavior-State (FBS) model by Umeda and Tomiyama et al.~\cite{umeda_1990, umeda_1996, umeda_2005}, which modeled function as a bridge between human intention and the physical behavior of an artifact.
While other researchers have approached the concept from different angles---such as Chandrasekaran and Josephson's~\cite{chandrasekaran_2000} focus on its meaning in engineering practice or Deng's~\cite{deng_2002} classification of functions---a common thread is the idea that function links the subjective world of human needs with the objective world of artifacts.
This is particularly salient in Balachandran and Gero's~\cite{balachandran_1991} argument that human subjectivity is an inextricable part of a function's definition.

Given our goal of connecting technology to consumer perception, we adopt the perspective of Umeda and Tomiyama, defining function as a subjective category that links human intentions and purposes to objective actions and structures.
For the practical description of these functions, we reviewed various methods, such as those discussed by Vermaas~\cite{vermaas_2013} and Keuneke~\cite{keuneke_1991}.
Among the organized representations summarized by Chakrabarti et al.~\cite{chakrabarti_1996}, we employ the ``verb-noun pair'' format:
\begin{itemize}
    \item Verb-noun pairs, such as ``transmit torque''; this representation has been around as long as humans have communicated with others how their designs do, or do not work.
    \item Input-output flow transformations, where the inputs and outputs can be energy, materials, or information
    \item Transformation between input-output situations
\end{itemize}

This format is simple, clear, and allows for a straightforward description of what a technology does, which can then be linked to the value a consumer feels.
This approach of treating functions as analyzable units is also consistent with recent work, such as the functional score proposed by Mun et al.~\cite{mun_2021}.

\subsection{Values: Understanding the Core of Consumer Needs}

The functions of a technology become meaningful only when they fulfill fundamental human values.
The ultimate success of a technology depends on whether consumers perceive it as valuable.
To systematically analyze this connection without being constrained by a company's preconceived notions, a comprehensive and structured theory of human values is required.

While early research like the Rokeach Value Survey~\cite{rokeach_1973} identified key values, it did not fully articulate the dynamic relationships between them.
A significant advancement was Schwartz's theory of basic individual values~\cite{schwartz_1992}.
Based on data from numerous countries, Schwartz proposed a theory of the universal content and structure of values, exploring the relationships of congruence and conflict among them.
The theory organizes values into ten broad motivational types (e.g., Hedonism, Security, Universalism) arranged in a circular continuum, as tested and visualized using Smallest Space Analysis (SSA).

\begin{figure}[ht]
    \centering
    \includegraphics[width=\linewidth]{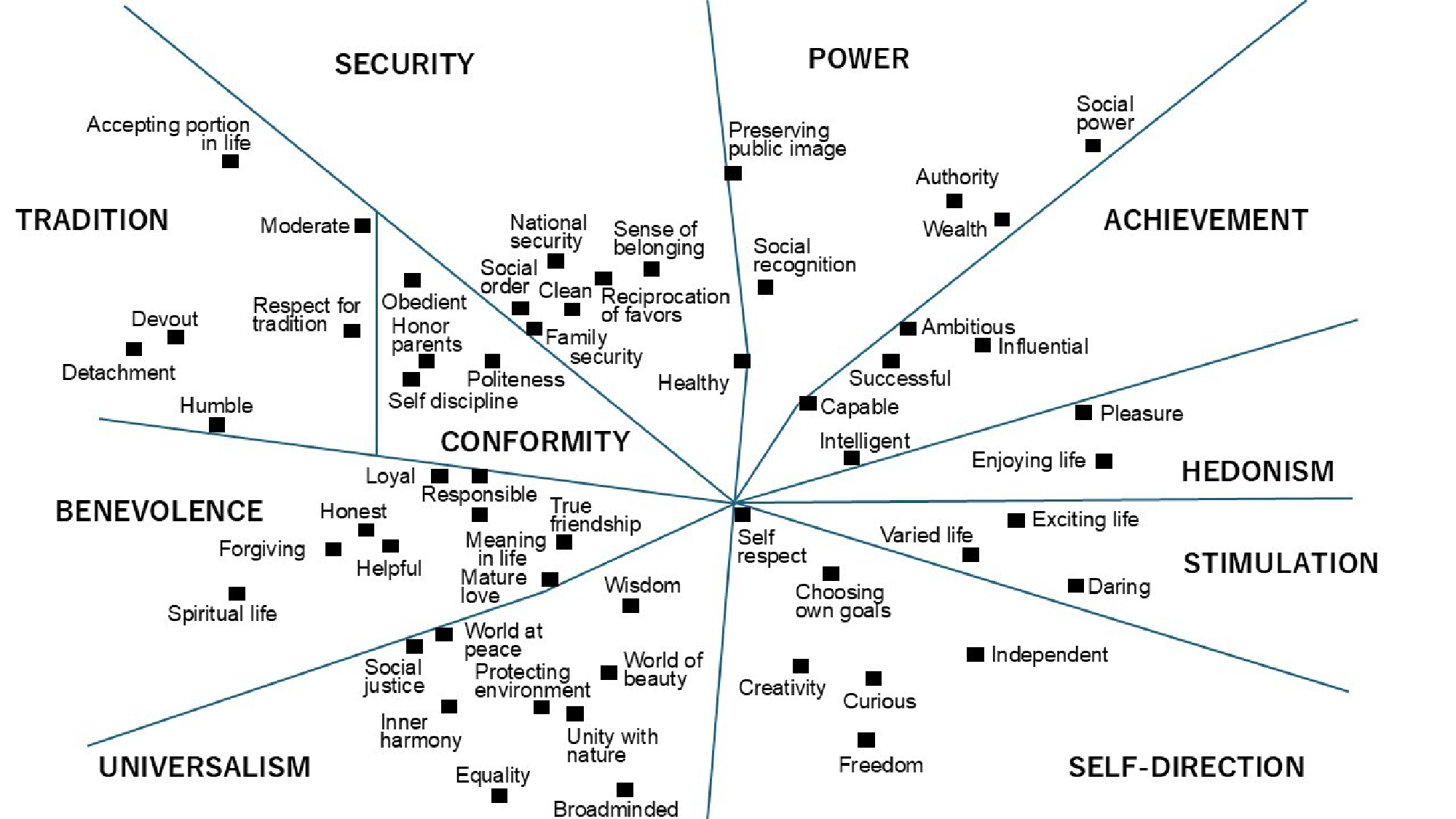}
    \caption{The pan-cultural model of human values derived from Schwartz's Smallest Space Analysis (SSA) of 36 samples across 20 countries. The plot illustrates how 56 single values form 10 motivationally distinct types, whose circular arrangement reflects a near-universal structure of conflict and compatibility. (Adapted from Figure 2 in Schwartz~\cite{schwartz_1992}).}
    \label{fig:value_structure}
\end{figure}

We chose Schwartz's theory as the foundation for our market-side evaluation for two primary reasons.
First, its comprehensiveness, validated extensively across cultures~\cite{schwartz_1994}, provides a robust framework to capture a wide range of potential consumer values, minimizing the risk of overlooking latent needs.
Second, its structural model of motivational compatibilities and conflicts is highly practical for our methodology.
For instance, if a technology strongly relates to one value type, the model suggests exploring adjacent, compatible value types for further opportunities.
This structural model is therefore particularly well-suited for the exploratory discovery process that our framework promotes.
A core principle of our methodology is to move beyond preconceived applications and systematically unearth a technology's latent value potential. Schwartz's structural model provides a conceptual map for this exploration.
It enables a logical expansion from an initial value perception to other, motivationally compatible values, thus supporting a discovery process that is both comprehensive and systematic, rather than reliant on serendipity.
While a refined 19-value theory was later proposed~\cite{schwartz_2012}, we utilize the original 10-value model as its broader categories are more suitable for initial brainstorming with consumers, who may not be accustomed to making fine-grained value distinctions.

\section{Methodology}\label{sec:method}
This section proposes a novel methodology to support Technology Opportunity Discovery (TOD).
The framework is specifically designed for situations where a company has developed core technologies in basic research and is exploring how to best utilize and commercialize them.
By systematically integrating two distinct perspectives—subjective value perception from the market and objective feasibility from technology experts—the methodology provides a structured foundation for supporting R\&D decision-making.

To ensure a clear and consistent basis for evaluation, each core technology was first decomposed into its constituent functions.
This critical preparatory step was conducted by three researchers based on the technical documentation provided by the company.
The researchers independently identified the core functionalities of each technology, describing them in the ``verb-noun pair'' format as discussed in Section~\ref{subsec:functions}.
Subsequently, they compared their lists and reached a consensus on the final set of functions to be evaluated in the workshops, thereby ensuring the objectivity and reproducibility of the process.

The proposed methodology consists of three consecutive steps: market-side evaluation, technology-side evaluation, and balancing.

\subsection{Market-side Evaluation}
As the first step, the methodology directly explores how consumers might use the technologies' defined functions and what value they might derive from them. 
The process requires the company's representatives (e.g., technology planners) to conduct a workshop with a group of potential customers (hereafter, ``participants''), who are asked to perform the following three tasks.

\subsubsection{Identification of Relevant Values}
First, participants are shown the developed technologies one by one and asked to select the most strongly related value-type for each technology from the 10 value-types defined by Schwartz~\cite{schwartz_1992}.
Because the $10$ value types can be too abstract for this task, the single values contained within each value type are also presented as a reference.

Next, for the selected value-type, a list of single values is generated, containing those from the value-type itself and its adjacent value types on the circular continuum.
This step is based on Schwartz's theory that adjacent values are motivationally compatible, allowing for a systematic exploration of related values.
For example, if \textit{Achievement} is selected, single values from \textit{Achievement}, \textit{Power}, and \textit{Hedonism} are listed for evaluation.
Participants then discuss and judge whether each single value on the list is related to the technology.

To facilitate this task, short additional explanations were created for each single value.
While Schwartz~\cite{schwartz_1992} provided brief descriptions, they were often too vague for participants to imagine a concrete relationship with technologies.
Therefore, the original explanations were carefully expanded to be more accessible.
The additional explanations and original descriptions are shown in \ref{app:exp_value}.

\subsubsection{Description of Potential Technology Usage Scenes}
This task enables the company to understand how consumers might actually use the technology.
For each single value identified in the previous task, participants are asked to describe a concrete usage scene where a consumer would feel that value through the utilization of the functions that the technology provides. Note that the functions are extracted beforehand as verb-noun pairs by experts who are well-versed in the technology within the company. 

As it can be difficult for a team to create a single scene from the outset, each individual is first asked to describe a scene for each single value.
Subsequently, the group discusses the individual scenes and collaboratively creates one unified scene that is considered to best represent the value.

\subsubsection{Scoring}
Through this task, all generated pairs of functions and single values (represented by scenes) are scored.
To anchor the scoring scale, participants are first asked to identify the single most desirable scene and the single least desirable scene from the entire list.
The most desirable scene is assigned a score of $7$, and the least desirable a score of $0$. Then, using these two scenes as benchmarks, participants score all other scenes on a relative scale from $0$ to $7$.

Finally, any pair of a function and a single value that did not result in a scene description is scored as $0$.

\subsection{Technology-side Evaluation}
In the second step, the methodology assesses the technical feasibility of the scenes desired by consumers.
The feasibility of each scene is evaluated by an internal expert (e.g., the lead engineer or a product manager intimately familiar with the technology's capabilities and limitations).
For this study, the evaluation was conducted by the chief of the department named ``technology deployment office'' at Sony CSL, who was selected for their unique position overseeing the entire lifecycle of the target technologies, from development to market implementation.
This expert possesses a comprehensive understanding of both the technologies' deep capabilities and the practical limitations of their commercialization, making their assessment highly reliable.

To focus the evaluation effort on the most promising ideas, only scenes that received a market-side score above a certain threshold (e.g., $5$ or higher in this study) are subjected to this technology-side evaluation.
This step ensures that technical feasibility assessment, which can be resource-intensive, is reserved for scenarios with demonstrated market appeal.

The evaluation uses a Technology Readiness Level (TRL) scale adapted from the one used in the EU's Horizon 2020 program, with additional explanations to make it suitable for assessing use-case scenarios, as shown in Table \ref{tab:TRLs_scene}.

{\rowcolors{5}{gray!15}{white}
\begin{longtable}{@{} c p{0.3\linewidth} p{0.5\linewidth} @{}}
    \caption{TRL scale adapted for evaluating the feasibility of use-case scenarios} \label{tab:TRLs_scene}\\
    \toprule 
    \textbf{Level} & \textbf{TRLs in the EU Horizon} & \textbf{Additional Explanation} \\
    \midrule 
    \endfirsthead
    
    \rowcolor{white} \caption[]{-- Continued} \\
    \toprule
    \rowcolor{white} \textbf{Level} & \textbf{TRLs in the EU Horizon} & \textbf{Additional Explanation} \\ 
    \midrule
    \endhead

    9 & Actual system proven in operational environment & 
    \begin{itemize}[leftmargin=*, topsep=0pt, partopsep=0pt, nosep]
        \item Stage where the system is actually implemented in the real world.
        \item Can be immediately implemented in the real world.
    \end{itemize} \\
    
    8 & System complete and qualified & 
    \begin{itemize}[leftmargin=*, topsep=0pt, partopsep=0pt, nosep]
        \item A system that can operate properly in the real world is constructed or can be constructed immediately.
        \item Although no system can be immediately implemented, other related technologies are prepared and can be constructed immediately.
    \end{itemize} \\

    7 & System prototype demonstration in operational environment & 
    \begin{itemize}[leftmargin=*, topsep=0pt, partopsep=0pt, nosep]
        \item A prototype system is built, and its operation in the real world has been demonstrated (or can be soon), although limited in scope.
    \end{itemize} \\

    6 & Technology demonstrated in relevant environment & 
    \begin{itemize}[leftmargin=*, topsep=0pt, partopsep=0pt, nosep]
        \item Demonstrated in a relevant field or could be demonstrated.
    \end{itemize} \\

    5 & Technology validated in relevant environment & 
    \begin{itemize}[leftmargin=*, topsep=0pt, partopsep=0pt, nosep]
        \item Validation of feasibility in a relevant field has been done or could be done soon.
    \end{itemize} \\

    4 & Technology validated in laboratory & 
    \begin{itemize}[leftmargin=*, topsep=0pt, partopsep=0pt, nosep]
        \item Validation of feasibility at a laboratory or simulation level has been done or could be done soon.
    \end{itemize} \\

    3 & Experimental proof of concept & 
    \begin{itemize}[leftmargin=*, topsep=0pt, partopsep=0pt, nosep]
        \item Concept has been demonstrated in the form of a laboratory or PoC (Proof of Concept).
    \end{itemize} \\

    2 & Technology concept formulated & 
    \begin{itemize}[leftmargin=*, topsep=0pt, partopsep=0pt, nosep]
        \item Basic concept is clarified or formulated.
        \item The technology can be adequately explained.
    \end{itemize} \\

    1 & Basic principles observed & 
    \begin{itemize}[leftmargin=*, topsep=0pt, partopsep=0pt, nosep]
        \item Stage where engineers understand the basic principles, methods, and theoretical certainty to some extent.
        \item Concepts are formed, and the basic principles to achieve them are understood to some extent.
    \end{itemize} \\
    \bottomrule
\end{longtable}
}

\subsection{Balancing}
In the final step, the methodology integrates the market-side and technology-side evaluations to identify TODs.

First, thresholds are set for both the market-side and technology-side scores.
These thresholds are not fixed; they should be determined by the company based on its strategic priorities, risk tolerance, and available resources.

Next, the scenes are filtered based on these thresholds.
Finally, scenes that surpass both the market and technology thresholds are identified as high-potential Technology Opportunities.
Each opportunity, therefore, consists of a concrete use-case scenario that links a specific technology function to a validated consumer value and is confirmed to be technologically feasible.
This resulting set of opportunities serves as a tangible input for subsequent strategic decision-making, such as prioritizing R\&D portfolios, shaping marketing messages, or developing new product concepts.
The two-dimensional evaluation (market value vs. technical feasibility) provides a powerful map for navigating the trade-offs between high-risk, high-return opportunities and more readily achievable ones.

\section{Experiment through workshops}\label{sec:workshop}
To test the validity and utility of our proposed methodology, we conducted a series of workshops in collaboration with Sony CSL.

\subsection{Experimental Design and Setup}
The experiment was designed not only to validate the methodology's overall process but also to explore how the perception of technology opportunities differs between potential end-users and technology experts.
To achieve this, we conducted workshops with two distinct participant groups: General Consumers and Sony CSL's Technology Deployment team.
By comparing the opportunities discovered by these two groups for the same set of technologies---which included both commercial successes and failures---we aim to identify patterns that may correlate with a technology's eventual market outcome.

Sony CSL, established in 1988 as a research institute of Sony Corporation, conducts a wide range of innovative research.
For our workshops, their Technology Deployment team, which is responsible for commercializing and implementing new technologies, selected four technologies they had developed: Omoiiro, Continuator, Cybercode, and Bubble Click.
Omoiiro and Cybercode were presented as examples of successful commercialization, while Continuator and Bubble Click were presented as unsuccessful cases.
This classification of ``success'' was determined by the Sony CSL Technology Deployment team based on their internal metrics, including market adoption and commercial viability.

In line with our methodology, each technology was first decomposed into its core functions.
As explained in Section \ref{subsec:functions}, this decomposition into independent functionalities, described using verb-noun pairs, is crucial for accurately capturing the relationship between what a technology does and the value consumers perceive.
The functions defined for each technology are shown in Table \ref{tab:functions_workshop}.

{\rowcolors{1}{white}{gray!15}
\begin{longtable}{@{} l p{0.6\linewidth} c @{}}
    \caption{Provided Technologies and Defined Functions for the Workshops} \label{tab:functions_workshop}\\
    \toprule 
    \textbf{Technology} & \textbf{Functions} & \textbf{Success}\\
    \midrule 
    \endfirsthead
    
    \caption[]{-- Continued} \\
    \toprule
    \rowcolor{white} \textbf{Technology} & \textbf{Functions} & \textbf{Success}\\
    \midrule
    \endhead

    Omoiiro & 
    \begin{enumerate}[leftmargin=*, topsep=0pt, partopsep=0pt, nosep]
        \item Learns the color scheme of an image and extracts the main colors that convey the atmosphere.
        \item Automatically creates a colorful color palette.
    \end{enumerate} 
    & $\checkmark$ \\
    
    Continuator & 
    \begin{enumerate}[leftmargin=*, topsep=0pt, partopsep=0pt, nosep]
        \item Extracts the characteristics of the music played by the user.
        \item Automatically generates continuous music in real-time.
    \end{enumerate} 
    & -- \\
    
    Cybercode & 
    \begin{enumerate}[leftmargin=*, topsep=0pt, partopsep=0pt, nosep]
        \item When a QR code is placed, it overlays CG in AR.
    \end{enumerate} 
    & $\checkmark$ \\
    
    Bubble Click & 
    \begin{enumerate}[leftmargin=*, topsep=0pt, partopsep=0pt, nosep]
        \item Calculates the position and popping of multiple bubbles, such as soap bubbles.
    \end{enumerate} 
    & -- \\
    \bottomrule
\end{longtable}}

\subsection{Market-side Evaluation Workshop}
The market-side evaluation workshops were conducted online on four dates in December 2022 and January 2023. 
Participants ($N=19$ total) consisted of both General Consumers ($N=9$) and members of the Technology Deployment team ($N=10$), with a broad age range from their 30s to 60s.
Each two-hour session was conducted online using a video conferencing application and the collaborative whiteboard platform Miro.
A facilitator from our research team was assigned to each group to ensure a smooth process.

Participants were divided into smaller groups, some composed entirely of General Consumers and others of the Technology Deployment team.
The specific composition of each workshop group, including the functions they evaluated, is detailed in Table~\ref{tab:workshop_settings}.
In the table, the ``Function'' column uses labels corresponding to the functions listed in Table \ref{tab:functions_workshop} (e.g., ``Continuator-1'' refers to the first function of Continuator).

\begingroup 
\renewcommand{\arraystretch}{1.3} 
\begin{longtable}{@{} lcccc @{}}
    \caption{Workshop Settings Summary} \label{tab:workshop_settings}\\
    \toprule
    \textbf{Date} & \textbf{Group Type} & \textbf{\# Participants} & \textbf{Function} & \textbf{Success}\\
    \midrule
    \endfirsthead
    
    \caption[]{-- Continued} \\
    \toprule
    \textbf{Date} & \textbf{Group Type} & \textbf{\# Participants} & \textbf{Functionality} & \textbf{Success}\\
    \midrule
    \endhead

    December 13, 2022 & General Consumers & 3 & Continuator-1 & -- \\
    December 13, 2022 & Technology Deployment & 3 & Omoiiro-1 & $\checkmark$ \\
    December 13, 2022 & Technology Deployment & 3 & Continuator-1 & -- \\
    \midrule 
    December 20, 2022 & General Consumers & 3 & Continuator-2 & -- \\
    December 20, 2022 & General Consumers & 3 & Cybercode & $\checkmark$ \\
    December 20, 2022 & Technology Deployment & 3 & Omoiiro-2 & $\checkmark$ \\
    December 20, 2022 & Technology Deployment & 3 & Bubble Click & -- \\
    December 20, 2022 & Technology Deployment & 4 & Continuator-2 & -- \\
    December 20, 2022 & Technology Deployment & 4 & Cybercode & $\checkmark$ \\
    \midrule 
    December 21, 2022 & General Consumers & 3 & Omoiiro-1 & $\checkmark$ \\
    December 21, 2022 & General Consumers & 3 & Omoiiro-2 & $\checkmark$ \\
    \midrule 
    January 6, 2023 & General Consumers & 3 & Bubble Click & -- \\
    \bottomrule
\end{longtable}
\endgroup 

The complete results of the market-side evaluation are provided in Appendix \ref{app:ws_market_results}.
As an illustrative example, Tables~\ref{tab:result_ws1} and \ref{tab:result_ws2} show the results for Omoiiro's first function from a workshop with General Consumers; ``Learns the color scheme of an image and extracts the main colors that convey the atmosphere.'' 

\begin{table}[h]
    \caption{An example result of the value identification step for Omoiiro.} 
    \label{tab:result_ws1}
    \begin{tabularx}{\textwidth}{@{} l X X @{}} 
        \toprule 
        \textbf{Selected Value Type} & \textbf{Related Single Values} & \textbf{Irrelevant Single Values}\\
        \midrule 
        Universalism & 
        Curious, Wisdom, Unity with nature, Broad minded, Equality, World of beauty, Helpful, Spiritual life & 
        Self respect, True friendship, Mature love, Independent, Honest, Meaning in life, Responsible, Choosing own goals, Forgiving, Loyal, Inner harmony, Freedom, Protecting the environment, Social justice, Creativity, World at peace \\
        \bottomrule
    \end{tabularx}
\end{table}

\begingroup
\rowcolors{1}{white}{gray!15}
\begin{longtable}{@{} l p{0.65\linewidth} c @{}}
    \caption{An example result of scene description and scoring for Omoiiro} \label{tab:result_ws2}\\
    \toprule 
    \textbf{Single Values} & \textbf{Scenes} & \textbf{Score}\\
    \midrule 
    \endfirsthead
    
    \caption[]{-- Continued} \\
    \toprule
    \rowcolor{white} \textbf{Single Values} & \textbf{Scenes} & \textbf{Score}\\
    \midrule
    \endhead

    Curious & I created a puzzle where you use the colors of a palette to make many small monochrome square pieces, and then recreate the original image using pointillism. & 5 \\ 
    Wisdom & I realized that the seemingly complex scenery we usually see is actually composed of a few simple colors, much like life can be simpler than it appears. & 2 \\
    Unity with nature & When abstracting both a photo of nature and my own painting, the same colors were extracted from both. & 3 \\
    Broad minded & Together with some friends, we each guessed the color we thought best represented the atmosphere of an image, discussed the reasons, and then compared our choices with the palette extracted by the program for feedback. & 5 \\
    Equality & By visualizing issues like educational inequality or gender disparities through color, it deepens understanding of the problems and inspires action toward solutions. & 6 \\ 
    World of beauty & I extracted colors from a photo of nature. The photo contained more colors than I expected, and seeing previously unrecognized colors revealed hidden beauty in nature. & 7 \\
    Helpful & I played a game with my child where we placed a palette next to an image and guessed where each color in the palette was used in the image. & 2 \\
    Spiritual life & The use of colors highlighted the environmental impact of mass production, making me more mindful of taking care of my belongings. & 0 \\
    \bottomrule
\end{longtable}
\endgroup

\subsection{Technology-side Feasibility Assessment}\label{subsec:trl_assessment}
Following the market-side evaluation, the feasibility of the generated use-case scenes was assessed.
This evaluation was performed by an experienced technology-deployment chief\footnote{%
The evaluator is one of the co-authors.
To mitigate potential bias,  
(i) the nine-level rubric in Table~\ref{tab:TRLs_scene} was finalised before the evaluator took part in any scoring;  
(ii) all scenes were randomised and anonymised so that their origin and market-side scores were hidden from the evaluator; and  
(iii) additional researchers, who were not involved in the initial scoring, reviewed the evaluator’s written justifications for each rating and confirmed their consistency with the rubric.%
}, who possesses deep knowledge of the technologies' capabilities and limitations.
In line with the methodology, and to focus analytical resources on the most promising scenarios, only scenes with a market-side score of $5$ or higher were selected for this TRL evaluation.
The expert then evaluated the TRL of each of these high-scoring scenes according to the scale presented in Table \ref{tab:TRLs_scene}.
The complete results are available in \ref{app:ws_trl_results}.
Table \ref{tab:result_ws_trl} provides an example of the TRL measurements for high-scoring scenes related to Omoiiro's first function, showing the scores assigned to scenes created by both ``General Consumers'' and the ``Technology Deployment'' team.

\begingroup
\rowcolors{1}{white}{gray!15}
\begin{longtable}{@{} l p{0.65\linewidth} c @{}}
    \caption{An example result of TRL measurement for Omoiiro} \label{tab:result_ws_trl}\\
    \toprule 
    \textbf{Single Values} & \textbf{Scenes} & \textbf{TRL}\\
    \midrule 
    \endfirsthead
    
    \caption[]{-- Continued} \\
    \toprule
    \rowcolor{white} \textbf{Single Values} & \textbf{Scenes} & \textbf{TRL}\\
    \midrule
    \endhead
    
    \rowcolor{white} 
    \multicolumn{3}{@{}c}{\textbf{Results from General Consumers Group}} \\
    \midrule
    Curious & I created a puzzle where you use the colors of a palette to make many small monochrome square pieces, and then recreate the original image using pointillism. & 6 \\
    Broad minded & Together with some friends, we each guessed the color we thought best represented the atmosphere of an image, discussed the reasons, and then compared our choices with the palette extracted by the program for feedback. & 6 \\
    Equality & By visualizing issues like educational inequality or gender disparities through color, it deepens understanding of the problems and inspires action toward solutions. & 1 \\
    World of beauty & I extracted colors from a photo of nature. The photo contained more colors than I expected, and seeing previously unrecognized colors revealed hidden beauty in nature. & 6 \\
    
    \midrule 
    
    \rowcolor{white} 
    \multicolumn{3}{@{}c}{\textbf{Results from Technology Deployment Group}} \\
    \midrule
    Creativity & Creators from different fields collaborate to create a piece (e.g., a painter and a ceramic artist), and the original colors can be extracted from the painter's work. & 9 \\
    Broad minded & Colors that go beyond implicit biases related to gender (e.g., warm or vivid tones for women's clothing) or habitual choices of designers can be extracted for original designs. & 9 \\
    Equality & Even people who lack confidence in their color sense can automatically extract colors just by selecting their favorite image. Non-designers may create more convincing color schemes than professional designers. & 6 \\
    Mature love & By using memorable photos, the colors based on those scenes or recognizable only by friends and family lead to a deeper attachment or familiarity with the work. & 6 \\
    Helpful & When out of design ideas, you can infinitely generate designs. & 9 \\
    Spiritual life & When there is an image supporting the design, a story emerges, making it more than just an object, but also providing spiritual support. & 6 \\
    \bottomrule
\end{longtable}
\endgroup

\section{Results}\label{sec:result}
The analysis of the workshop data reveals two distinct patterns that appear to differentiate the commercially successful technologies from the unsuccessful ones.
We term these patterns the ``\textit{vision gap}'' and ``\textit{value breadth}.''
This section presents these findings, supported by visualizations and quantitative comparisons.

\begin{figure}[htbp]
    \centering 
    
    \begin{subfigure}[b]{0.48\textwidth}
        \centering
        \includegraphics[width=\linewidth]{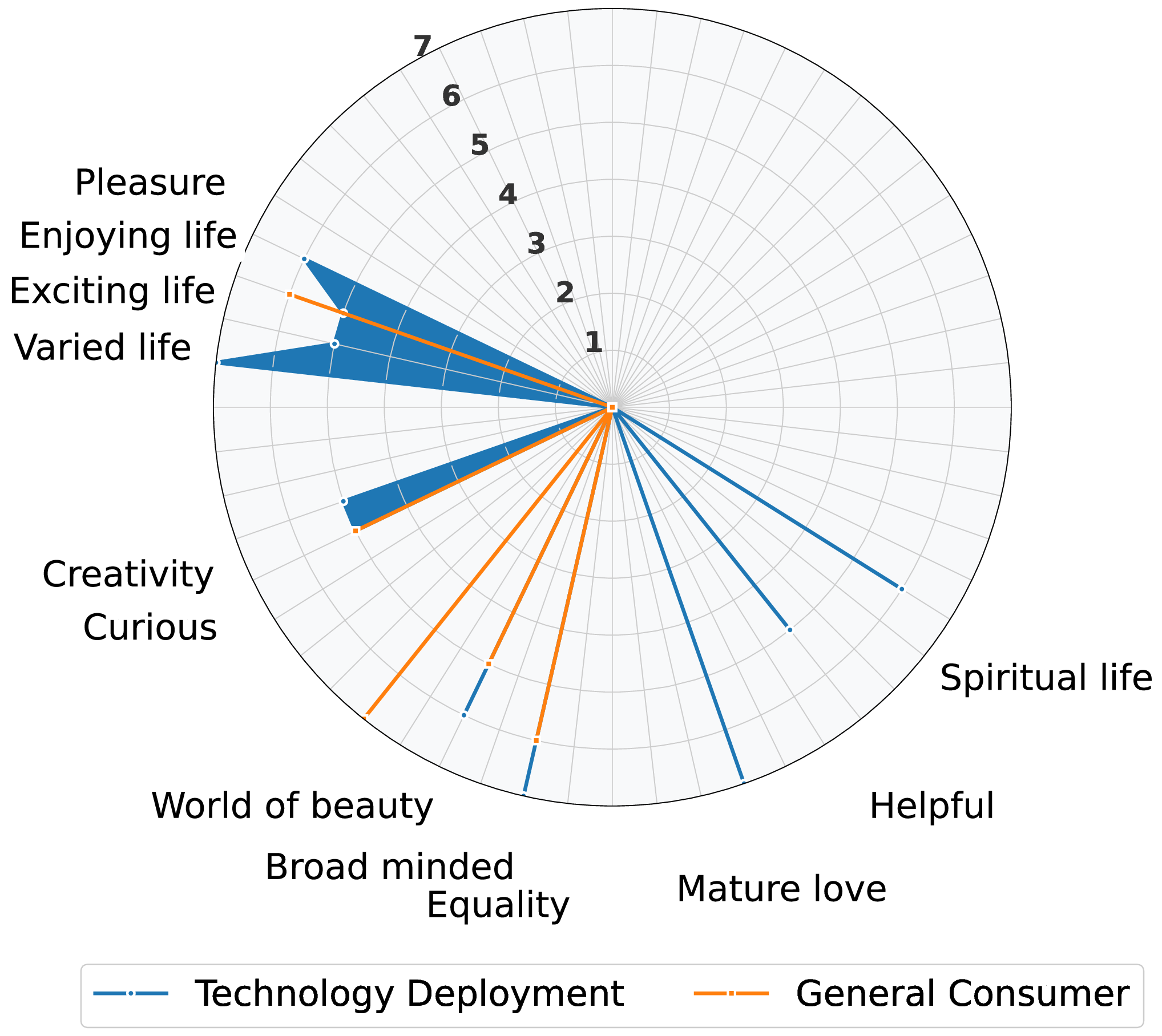}
        \caption{TOD found for ``Omoiiro'' (Success)}
        \label{fig:TOD_omoiiro}
    \end{subfigure}
    \hfill 
    \begin{subfigure}[b]{0.48\textwidth}
        \centering
        \includegraphics[width=\linewidth]{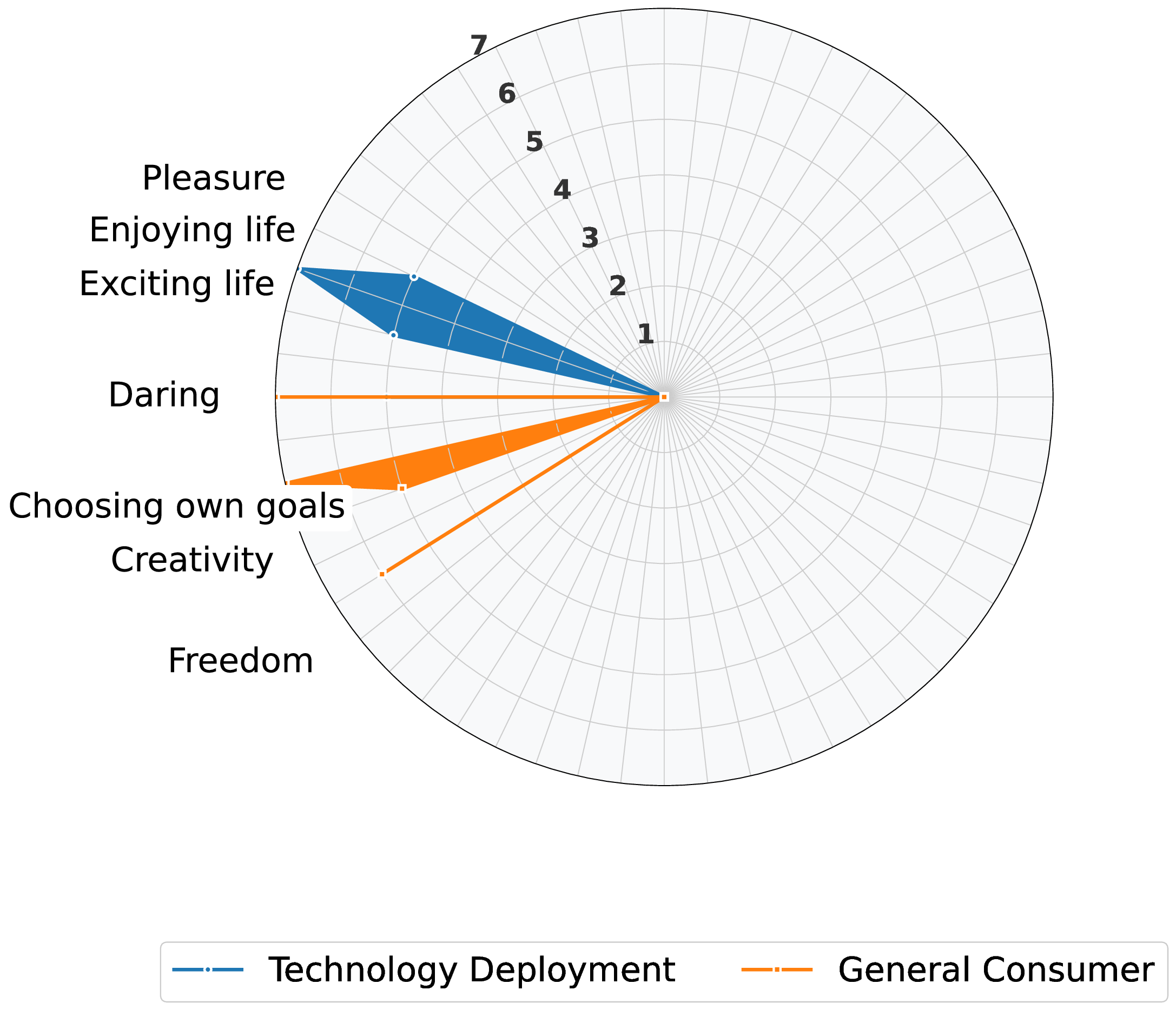}
        \caption{TOD found for ``Cybercode'' (Success)}
        \label{fig:TOD_cybercode}
    \end{subfigure}

    \vspace{1cm} 

    \begin{subfigure}[b]{0.48\textwidth}
        \centering
        \includegraphics[width=\linewidth]{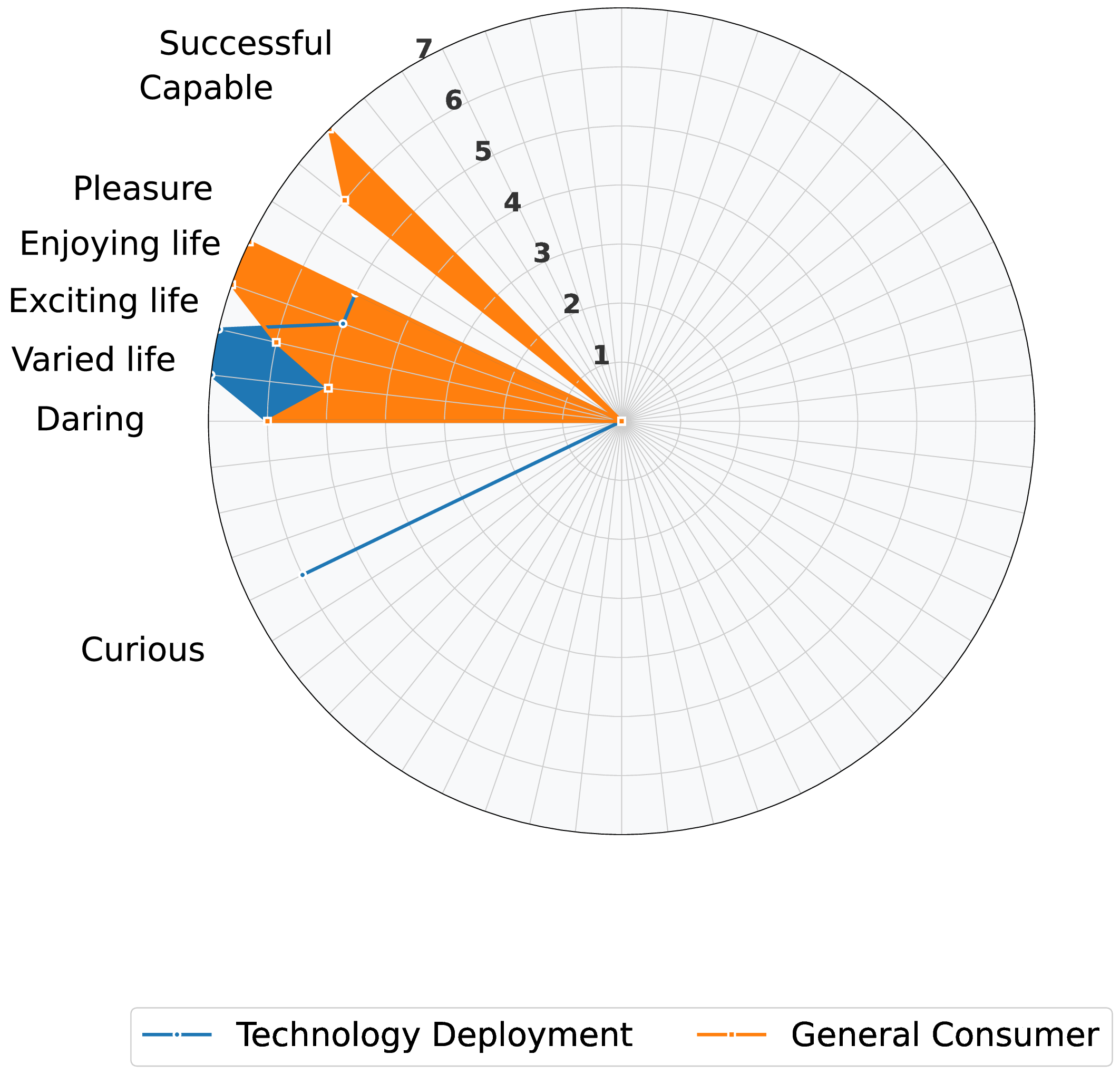}
        \caption{TOD found for ``Continuator'' (Failure)}
        \label{fig:TOD_continuator}
    \end{subfigure}
    \hfill 
    \begin{subfigure}[b]{0.48\textwidth}
        \centering
        \includegraphics[width=\linewidth]{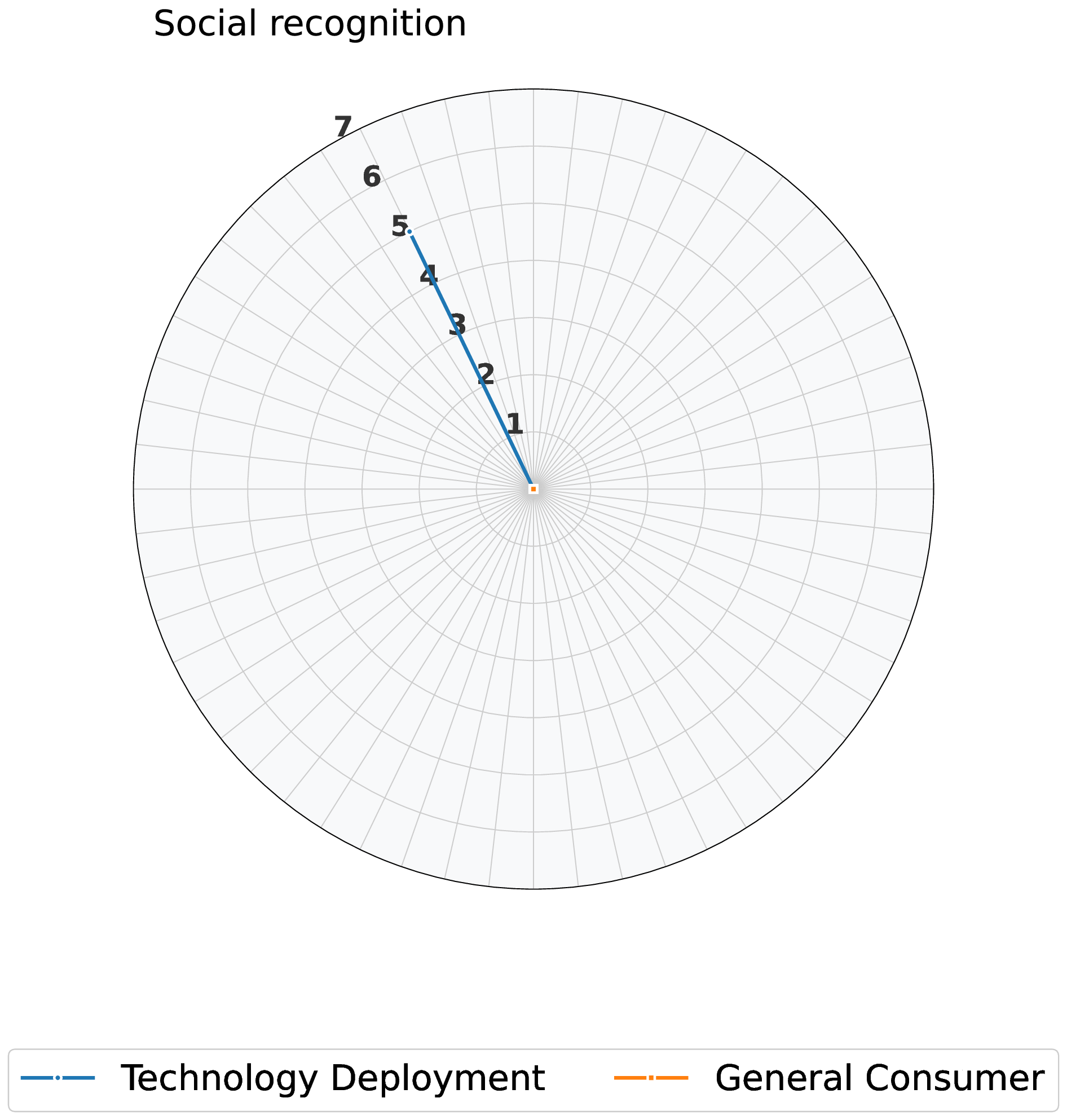}
        \caption{TOD found for ``Bubble Click'' (Failure)}
        \label{fig:TOD_bubbleclick}
    \end{subfigure}
    
    \caption{TODs found for the four technologies. The results for successful cases (a, b) and failure cases (c, d) are compared. The blue line represents the Technology Deployment team's evaluation, and the orange line represents the General Consumers' evaluation. Note that for ``Bubble Click,'' no opportunities identified by General Consumers met the specified thresholds.}
    \label{fig:TOD_all_results} 
\end{figure}

The primary output of our methodology is a set of high-potential technology opportunities for each technology, visualized as radar charts in Figure~\ref{fig:TOD_all_results}.
These charts illustrate the TODs identified through the workshops.
Each axis corresponds to a single value from Schwartz's theory, and the plotted points represent the score (from $0$ to $7$) of the use-case scenes associated with that value.
In accordance with the methodology, only opportunities that surpassed the thresholds for both market-side evaluation (scene score $\geq 5$) and technology-side evaluation (TRL $\geq 5$) are plotted.
The blue lines represent the opportunities identified by the Technology Deployment team, while the orange lines represent those identified by General Consumers.
Where multiple functions of a technology were linked to the same single value, only the opportunity with the highest market-side score is shown.

Our first key finding is the ``\textit{vision gap}'': a qualitative disparity in the scope of perceived values between internal experts (Technology Deployment team) and General Consumers.
As summarized in Table~\ref{tab:value_type_comparison}, this pattern is particularly evident when comparing the successful case (Omoiiro) and the failure case (Continuator).
For the successful technology, Omoiiro, the value types identified by the Technology Deployment team fully encompassed those identified by consumers.
This inclusive pattern was not observed for the unsuccessful technology, Continuator, where the value types identified by the two groups only partially overlapped, suggesting a potential misalignment in vision.

\begin{table}[h]
    \centering
    \caption{Comparison of Value Types Identified by Technology Deployment and Consumers}
    \label{tab:value_type_comparison}
    \begin{tabularx}{0.9\textwidth}{@{} l X X @{}}
        \toprule 
        & \textbf{Omoiiro} & \textbf{Continuator} \\
        \midrule
        \textbf{Value Types (Tech Deployment)} & 
        Hedonism, Stimulation, Self-direction, Universalism, Benevolence & 
        Self-direction, Hedonism, Stimulation \\
        
        \addlinespace 

        \textbf{Value Types (Consumers)} & 
        Universalism, Self-direction, Hedonism & 
        Stimulation, Hedonism, Achievement \\
        \bottomrule
    \end{tabularx}
\end{table}

Our second key finding concerns ``\textit{value breadth},'' which we define as the number of distinct, high-potential single values associated with a technology from the experts' perspective.
A quantitative analysis reveals a clear disparity between successful and unsuccessful cases.
The successful technologies, Omoiiro and Cybercode, were associated with $11$ and $4$ high-potential values, respectively.
In contrast, the unsuccessful technologies, Continuator and Bubble Click, were associated with only $6$ and $1$ values.
This trend holds even when comparing technologies with the same number of functions: the successful Omoiiro ($11$ values) demonstrated a significantly greater \textit{value breadth} than the unsuccessful Continuator ($6$ values), suggesting that a technology's versatility in evoking diverse values may be an indicator of its market potential.

\section{Discussion}\label{sec:discussion}
The results present two main findings—the ``\textit{vision gap}'' and ``\textit{value breadth}''—that appear to distinguish successful from unsuccessful technologies.
Here, we interpret these findings and discuss their implications for both management practice and innovation theory.

\subsection{The ``\textit{Vision Gap}'' as a Driver of Success}
The first finding suggests that for a technology to succeed, its internal technology deployment team must possess a vision for its potential that is broader than the consumers' initial perceptions.
In the case of the successful technology Omoiiro, the experts not only understood the values consumers expect (e.g., Hedonism, Self-direction) but also identified additional value dimensions (Stimulation, Benevolence).
We term this the ``\textit{vision gap}.''
This gap appears crucial for guiding marketing, product development, and strategic positioning to unlock a technology's full market potential, moving beyond its most obvious applications.

This directly addresses the limitation of purely market-driven approaches identified in the introduction.
Relying solely on initial consumer feedback might have constrained Omoiiro's development to a narrower set of applications, whereas the experts' broader vision could have paved the way for its success.

Furthermore, the \textit{vision gap} may not only be a matter of breadth but also of alignment.
In the case of the unsuccessful technology, Continuator, a misalignment in the direction of values was observed.
The experts primarily envisioned values in the Self-direction domain, whereas consumers sought values related to Achievement.
This suggests that a successful vision must not only be broader but also build upon or complement the values perceived by consumers.
A vision that is completely disconnected from consumer perception risks being a ``mismatch'' rather than a strategic ``gap,'' ultimately failing to gain market traction.

\subsection{``\textit{Value Breadth}'' as an Indicator of Potential}
The second finding suggests that ``\textit{value breadth}''---the number of distinct values a technology can evoke---may serve as a lead indicator of its commercial potential. 
A technology that resonates with a wider array of human values is likely more versatile and adaptable.
It can appeal to a broader customer base, find application in more diverse contexts, and prove more resilient to shifts in market preferences.
While this analysis is based on a limited set of cases, the clear difference in \textit{value breadth} between the successful ($11$ and $4$ values) and unsuccessful ($6$ and $1$ values) technologies is a compelling pattern.

This metric offers a new, human-centric lens for early-stage technology assessment.
While TRL measures technical maturity, \textit{value breadth} gauges market versatility.
A high \textit{value breadth} score suggests a technology has the potential to evolve beyond a single niche application and create value across multiple market segments, some of which may be unforeseen at the outset.

\subsection{Implications for R\&D Management}
For practitioners in technology management, our study offers a concrete methodology for systematically mapping technologies to consumer values.
The concepts of the ``\textit{vision gap}'' and ``\textit{value breadth}'' can serve as novel diagnostic tools.
Managers can use this framework to address critical questions:
\begin{itemize}
    \item Is our internal vision aligned with and sufficiently broader than the market's perception? (\textit{vision gap} analysis)
    \item How versatile is this technology in terms of the human needs it can satisfy? (\textit{value breadth} analysis)
\end{itemize}
By providing data-driven answers to these questions, our methodology helps to replace intuition with evidence in the fuzzy front-end of innovation, enabling a more objective and systematic prioritization of technologies within an R\&D portfolio.

\subsection{Theoretical Implications}
For theory, this study provides a novel application of Schwartz's value theory, operationalizing it as a decision support tool in technology management.

It offers an empirical bridge between the abstract, psychological structure of human values and the tangible outcomes of technology commercialization.
This work pioneers a new path for integrating human-centric psychological theories into innovation management research, responding to the call for methodologies that can systematically incorporate end-user values into early-stage technological planning.

\section{Conclusion}\label{sec:conclusion}

This paper addressed the critical challenge of technology opportunity discovery (TOD), particularly the difficulty of integrating consumer perspectives in the early stages of technology development. 
We proposed and validated a novel framework that bridges this gap by integrating a technology-centric view (Technology Readiness Levels) with a human-centric one (the Schwartz theory of basic values). 
The methodology, implemented through a series of workshops with Sony CSL, operationalizes this integration through a three-step process: market-side evaluation, technology-side evaluation, and balancing. 
Our primary contribution lies in the empirical findings derived from this framework. 
The analysis of successful and unsuccessful technologies revealed two key concepts: the ``\textit{vision gap},'' where the development team's broader perception of a technology's value potential compared to consumers' initial views correlates with success; and ``\textit{value breadth},'' the number of values a technology evokes, which appears to be an indicator of its market potential.

Despite these contributions, this study has several limitations that open avenues for future research. We recognize these as crucial steps toward refining and generalizing the proposed framework.
\begin{itemize}
    \item \textbf{Subjectivity and Constraints in the Elicitation Process:}
    The workshop process itself has limitations.
    The extraction of functions was based on researcher interpretation of descriptive texts, introducing potential subjectivity.
    Furthermore, the value elicitation method, while practical, was constrained to a subset of values based on the initial selection, risking the omission of other relevant values.
    The group discussion format is also subject to facilitator bias and group dynamics. Future work should aim to establish more objective protocols for these steps.

    \item \textbf{Scalability and Data Scale:}
    The current methodology, relying on in-depth workshops and expert interviews, is resource-intensive.
    The small sample size (four technologies, $N=19$ participants) limits the statistical power of the findings.
    Similarly, relying on a single key expert for technical assessment is not easily scalable.
    Future research should explore more efficient and scalable methods, such as quantitative surveys for market data and patent analysis for technical assessment.
    
    \item \textbf{Realism of the Matching Solution:}
    The framework identifies all opportunities that surpass a simple threshold.
    This does not account for real-world business constraints, such as limited R\&D budgets.
    Future work could incorporate such constraints and employ formal matching or optimization algorithms to generate more realistic and actionable sets of TODs.
\end{itemize}

Acknowledging these limitations, this study nonetheless carves a novel and promising path toward a more human-centric and evidence-informed approach to technology opportunity discovery.
We believe the proposed framework and the concepts derived from it offer a valuable foundation for both academics and practitioners seeking to navigate the complex interface between technology and society.

\section*{Appendix A. Additional Explanation of single values}
\label{app:exp_value}

This appendix lists the $56$ single values based on the Schwartz Value Survey, which were used in our workshops.
The ``Original Description'' column shows the explanatory phrase provided in Schwartz~\cite{schwartz_1992}.
The ``Additional Explanation'' column shows the revised descriptions created for this study to enhance clarity and ensure a common understanding among workshop participants, as mentioned in the Methodology section.

\begingroup 
\footnotesize 
\rowcolors{2}{white}{gray!15}
\begin{longtable}{@{} >{\raggedright\arraybackslash}p{0.15\linewidth}>{\raggedright\arraybackslash}p{0.15\linewidth}>{\raggedright\arraybackslash}p{0.3\linewidth}>{\raggedright\arraybackslash}p{0.3\linewidth} @{}}
    \caption{List of Single Values with Original and Additional Explanations} \label{tab:exp_value}\\
    \hline
    \rowcolor{white} 
    \textbf{Value Type} & \textbf{Single Value} & \textbf{Original Description} & \textbf{Additional Explanation}\\
    \hline 
    \endfirsthead
    
    \caption[]{-- Continued} \\
    \hline
    \rowcolor{white} 
    \textbf{Value Type} & \textbf{Single Value} & \textbf{Original Description} & \textbf{Additional Explanation}\\
    \hline
    \endhead
    \hline
    \endlastfoot

    Power & Social power & control over others, dominance & Feeling dominant or superior over others \\
    \cline{2-4}
    & Wealth & material possessions, money & Feeling materially and financially fulfilled \\
    \cline{2-4}
    & Authority & the right to lead or command & Feeling a sense of leadership \\
    \cline{2-4}
    & Preserving my public image & protecting my face & Being able to maintain one's face or reputation \\
    \cline{2-4}
    & Social recognition & respect, approval by others & Feeling respected and recognized by others \\
    \hline
    
    Achievement & Influential & having an impact on people and events & Being able to influence people or events \\
    \cline{2-4}
    & Ambitious & hardworking, aspiring & Striving to improve oneself \\
    \cline{2-4}
    & Successful & achieving goals & Feeling a sense of achievement \\
    \cline{2-4}
    & Capable & competent, effective, efficient & Becoming effective and efficient \\
    \cline{2-4}
    & Intelligent & logical, thinking & Leading to logical thinking \\
    \hline
    
    Hedonism & Pleasure & gratification of desires & Feeling the joy of having desires fulfilled \\
    \cline{2-4}
    & Enjoying life & enjoying food, sex, leisure, etc. & Gaining enjoyment from life \\
    \hline

    
    Stimulation & Exciting life & stimulating experiences & Experiencing exciting activities \\
    \cline{2-4}
    & Varied life & filled with challenge, novelty and change & Feeling change and novelty \\
    \cline{2-4}
    & Daring & seeking adventure, risk & Encouraging adventure and risk-taking \\
    \hline
    
    Self-direction & Self Respect & belief in one’s own worth & Being able to be oneself \\
    \cline{2-4}
    & Choosing own goals & selecting own purposes & Leading to the selection of one's own goals \\
    \cline{2-4}
    & Creativity & uniqueness, imagination & Feeling unique and creative \\
    \cline{2-4}
    & Curious & interested in everything, exploring & Stimulating curiosity and exploration \\
    \cline{2-4}
    & Freedom & freedom of action and thought & Feeling free to act and think \\
    \cline{2-4}
    & Independent & self-reliant, self-sufficient & Being self-sufficient and not relying on others \\
    \hline

    Universalism & Wisdom & a mature understanding of life & Gaining a deep understanding of life \\
    \cline{2-4}
    & World of beauty & beauty of nature and the arts & Appreciating the beauty of nature and the arts \\
    \cline{2-4}
    & Unity with nature & fitting into nature & Feeling at one with nature \\
    \cline{2-4}
    & Broad minded & tolerant of different ideas and beliefs & Becoming tolerant of different ideas and beliefs \\
    \cline{2-4}
    & Protecting the environment & preserving nature & Contributing to environmental preservation \\
    \cline{2-4}
    & Equality & equal opportunity for all & Providing equal opportunities for all \\
    \cline{2-4}
    & World at peace & free of war and conflict & Bringing world peace \\
    \cline{2-4}
    & Social justice & correcting injustice, care for the weak & Contributing to correcting injustice and protecting the weak \\
    \cline{2-4}
    & Inner harmony & at peace with myself & Feeling inner peace \\
    \hline

    Benevolence & True friendship & close, supportive friends & Feeling closeness like friendship \\
    \cline{2-4}
    & Mature love & deep emotional and spiritual intimacy & Feeling affection and love \\
    \cline{2-4}
    & Meaning in life & a purpose in life & Feeling a sense of purpose in life \\
    \cline{2-4}
    & Responsible & dependable, reliable & Feeling dependable \\
    \cline{2-4}
    & Helpful & working for the welfare of others & Being able to help others \\
    \cline{2-4}
    & Honest & genuine, sincere & Feeling honest and sincere \\
    \cline{2-4}
    & Forgiving & willing to pardon others & Feeling forgiving \\
    \cline{2-4}
    & Spiritual life & emphasis on spiritual not material matters & Focusing on spiritual rather than material fulfillment \\
    \cline{2-4}
    & Loyal & faithful to my friends, group & Being loyal to friends and groups \\
    \hline

    Conformity & Politeness & courtesy, good manners & Being polite and having good manners \\
    \cline{2-4}
    & Honoring of parents and elders & showing respect & Showing respect to parents and elders \\
    \cline{2-4}
    & Self discipline & self-restraint, resistance to temptation & Maintaining self-discipline and resisting temptation \\
    \cline{2-4}
    & Obedient & dutiful, meeting obligations & Being dutiful and fulfilling obligations \\
    \hline

    Tradition & Respect for tradition & preservation of time-honored customs & Contributing to the preservation of tradition \\
    \cline{2-4}
    & Devout & holding to religious faith and belief & Maintaining religious faith \\
    \cline{2-4}
    & Detachment & from worldly concerns & Gaining detachment from worldly concerns \\
    \cline{2-4}
    & Humble & modest, self-effacing & Feeling humility and modesty \\
    \cline{2-4}
    & Moderate & avoiding extremes of feeling and action & Feeling balanced without extremes in emotion or action \\
    \cline{2-4}
    & Accepting portion in life & submitting to life’s circumstances & Being able to adapt and accept life’s circumstances \\
    \hline

    Security & Healthy & not being sick physically or mentally & Promoting physical and mental health \\
    \cline{2-4}
    & Sense of belonging & feelings that others care about me & Feeling a sense of belonging and care from others \\
    \cline{2-4}
    & National security & protection of my nation from enemies & Contributing to national security \\
    \cline{2-4}
    & Social order & stability of society & Contributing to social stability and order \\
    \cline{2-4}
    & Family security & safety for loved ones & Contributing to the safety of loved ones \\
    \cline{2-4}
    & Reciprocation of favor & avoidance of indebtedness & Feeling reciprocal and mutually beneficial \\
    \cline{2-4}
    & Clean & neat, tidy & Feeling cleanliness and neatness, maintaining cleanliness \\

\end{longtable}
\endgroup

\section*{Appendix B. Detailed results of workshops: market-side evaluation}\label{app:ws_market_results}

This appendix provides the detailed results from the market-side evaluation workshops, as described in the Methodology and Experiment sections. Table \ref{tab:value_identification_combined} presents the results from the value identification step.
Subsequently, Tables \ref{tab:scenes_consumers} and \ref{tab:scenes_deployment} present the detailed use-case scenes and their corresponding scores.
Within each table, the results from the ``General Consumers'' and ``Technology Deployment'' groups are presented for comparison.

\begingroup
\footnotesize 
\rowcolors{2}{white}{gray!15}
\begin{longtable}{@{} >{\raggedright\arraybackslash}p{0.1\linewidth} >{\raggedright\arraybackslash}p{0.1\linewidth} >{\raggedright\arraybackslash}p{0.1\linewidth} >{\raggedright\arraybackslash}p{0.3\linewidth} >{\raggedright\arraybackslash}p{0.3\linewidth} @{}}
    \caption{The result of identification of values by participant group.}
    \label{tab:value_identification_combined}\\
    \toprule
    \rowcolor{white} 
    \textbf{Group} & \textbf{Function} & \textbf{Selected Value Type} & \textbf{Related Single Values} & \textbf{Irrelevant Single Values} \\
    \midrule
    \endfirsthead
    
    \caption[]{-- Continued} \\
    \toprule
    \rowcolor{white} 
    \textbf{Group} & \textbf{Function} & \textbf{Selected Value Type} & \textbf{Related Single Values} & \textbf{Irrelevant Single Values} \\
    \midrule
    \endhead
    \bottomrule
    \endlastfoot

    General Consumers & Omoiiro-1 & Universalism & Curious, Wisdom, Unity with nature, Broad minded, Equality, World of beauty, Helpful, Spiritual life & Self respect, True friendship, Mature love, Independent, honest, Meaning in life, Responsible, Choosing own goals, Forgiving, Loyal, Inner harmony, Freedom, Protecting the environment, Social justice, Creativity, World at peace \\
    Technology Deployment & Omoiiro-1 & Universalism & Self Respect, Creativity, Curious, Freedom, World of beauty, Unity with nature, Broad minded, Equality, Inner harmony, True friendship, Mature love, Helpful, Spiritual life & Choosing own goals, Independent, Wisdom, Protecting the environment, World at peace, Social justice, Meaning in life, Responsible, Honest, Forgiving, Loyal \\

    General Consumers & Omoiiro-2 & Hedonism & Influential, Ambitious, Capable, Enjoying life, Exciting life, Varied life & Successful, Pleasure, Daring, Intelligent \\
    Technology Deployment & Omoiiro-2 & Stimulation & Pleasure, Enjoying life, Exciting life, Varied life, Creativity, Curious, Freedom, Self Respect, Choosing own goals, Independent & Daring \\

    General Consumers & Continuator-1 & Hedonism & Influential, Ambitious, Successful, Capable, Pleasure, Enjoying Life, Exciting Life, Varied life, Daring & Intelligent \\
    Technology Deployment & Continuator-1 & Stimulation & Pleasure, Enjoying life, Exciting life, Varied life, Self Respect, Choosing own goals, Creativity, Curious, Freedom, Independent & Daring \\

    General Consumers & Continuator-2 & Hedonism & Influential, Ambitious, Successful, Capable, Pleasure, Enjoying life, Exciting life, Varied life, Daring & Intelligent \\
    Technology Deployment & Continuator-2 & Stimulation & Pleasure, Enjoying life, Exciting life, Varied life, Daring, Self Respect, Creativity, Curious, Independent & Choosing own goals, Freedom \\

    General Consumers & Cybercode-1 & Stimulation & Pleasure, Enjoying life, Exciting life, Varied life, Daring, Choosing own goals, Creativity, Curious, Freedom & Self respect, Independent \\
    Technology Deployment & Cybercode-1 & Stimulation & Pleasure, Enjoying life, Exciting life, Varied life, Daring, Curious, Freedom & Self respect, Choosing own goals, Creativity \\
    
    General Consumers & Bubble Click-1 & Hedonism & Influential, Successful, Pleasure, Enjoying life, Exciting life & Ambitious, Capable, Intelligent, Varied life, Daring \\
    Technology Deployment & Bubble Click-1 & Achievement & Social recognition, Influential, Successful, Capable, Enjoying life & Intelligent, Social Power, Wealth, Preserving my public image, Authority, Pleasure, Ambitious \\
\end{longtable}
\endgroup

\begingroup
\footnotesize 
\centering
\rowcolors{2}{white}{gray!15}
\begin{longtable}{>{\raggedright\arraybackslash}p{0.15\textwidth} >{\raggedright\arraybackslash}p{0.13\textwidth} >{\raggedright\arraybackslash}p{0.5\linewidth} c}
    \caption{Result of scene description and scoring by General Consumers.} \label{tab:scenes_consumers}\\
    \hline
    \rowcolor{white} 
    \textbf{Function} & \textbf{Single Values} & \textbf{Scenes} & \textbf{Score}\\
    \hline
    \endfirsthead
    
    \caption[]{-- Continued} \\
    \hline
    \rowcolor{white} 
    \textbf{Function} & \textbf{Single Values} & \textbf{Scenes} & \textbf{Score}\\
    \hline
    \endhead
    \hline
    \endlastfoot

    Omoiiro-1 & Curious & I created a puzzle where you use the colors of a palette to make many small monochrome square pieces, and then recreate the original image using pointillism. & 5 \\ 
    \cline{2-4}
    & Wisdom & I realized that the seemingly complex scenery we usually see is actually composed of a few simple colors, much like life can be simpler than it appears. & 2 \\ 
    \cline{2-4}
    & Unity with nature & When abstracting both a photo of nature and my own painting, the same colors were extracted from both. & 3 \\ 
    \cline{2-4}
    & Broad minded & Together with some friends, we each guessed the color we thought best represented the atmosphere of an image, discussed the reasons, and then compared our choices with the palette extracted by the program for feedback. & 5 \\ 
    \cline{2-4}
    & Equality & By visualizing issues like educational inequality or gender disparities through color, it deepens understanding of the problems and inspires action toward solutions. & 6 \\ 
    \cline{2-4}
    & World of beauty & I extracted colors from a photo of nature. The photo contained more colors than I expected, and seeing previously unrecognized colors revealed hidden beauty in nature. & 7 \\
    \cline{2-4}
    & Helpful & I played a game with my child where we placed a palette next to an image and guessed where each color in the palette was used in the image. & 2 \\ 
    \cline{2-4}
    & Spiritual life & The use of colors highlighted the environmental impact of mass production, making me more mindful of taking care of my belongings. & 0 \\
    \hline

    Omoiiro-2 & Ambitious & By guessing the colors used in a photo and then outputting them as a color palette, you can improve your color recognition ability while striving to think of combinations beyond those presented. & 3 \\ 
    \cline{2-4}
    & Capable & With recommended color combinations, anyone can coordinate outfits, interior design, or even the colors in PPTs or research graphs at a certain level. & 4 \\ 
    \cline{2-4}
    & Enjoying life & By converting a photo you took into a palette, you can express your current feelings, share it on social media, and feel a sense of connection if someone else has a similar color palette. & 6 \\ 
    \cline{2-4}
    & Exciting life & Input your mood for the day, and based on the recommended palette, you can choose clothes that reflect that mood, leading to an exciting day as you go out wearing them. & 7 \\ 
    \cline{2-4}
    & Varied life & You can more clearly feel the changing of the seasons through the shift in color palettes, and reflect this in your hair color or clothing coordination. & 6 \\
    \hline

    Continuator-1 & Influential & A person who wants to compose music can use this to receive suggestions for new songs. They can gain new inspiration and convey their own thoughts. & 0 \\
    \cline{2-4}
    & Ambitious & A person who wants to compose music can arrive at their preferred style or characteristics while altering certain features. & 4 \\
    \cline{2-4}
    & Successful & A person who wants to compose music can create their desired song and feel a sense of accomplishment. & 7 \\
    \cline{2-4}
    & Capable & A person who wants to compose music can efficiently move closer to their desired song by extracting characteristics from hit songs, among other things. & 6 \\
    \cline{2-4}
    & Pleasure & Even if someone wants to compose music but cannot do it well, they can compose just by inputting their desired song image at the beginning. They can feel joy as their wish to compose music is fulfilled. & 6 \\
    \cline{2-4}
    & Enjoying Life & A person who wants to compose music can add a new enjoyment to their life by using this. & 2 \\
    \cline{2-4}
    & Exciting Life & General consumers can have an exciting experience by creating their own music. & 6 \\
    \cline{2-4}
    & Varied Life & A person who wants to compose music can create different types of songs without repeating similar phrases. They can create songs with different features and new images. & 5 \\
    \cline{2-4}
    & Daring & A person who wants to compose music can explore and venture into new genres and styles that differ from their own. & 4 \\
    \hline

    Continuator-2 & Influential & The phrases or music you create are recognized by others. & 0 \\ 
    \cline{2-4}
    & Ambitious & Even those with no experience can compose music, and they can revise and remake their compositions multiple times. & 4 \\ 
    \cline{2-4}
    & Successful & You can generate the music you desire, including long pieces, without much effort, giving you a sense of accomplishment. & 3 \\ 
    \cline{2-4}
    & Capable & You can compose music in a short amount of time, and create songs efficiently, regardless of your level of learning or experience. & 2 \\ 
    \cline{2-4}
    & Pleasure & You feel joy in composing music, even without any skills or prior experience. & 7 \\ 
    \cline{2-4}
    & Enjoying life & You find enjoyment in the ease of composing music as part of your everyday life. & 7 \\ 
    \cline{2-4}
    & Exciting life & Unexpected songs may emerge or be performed, making it a stimulating experience. & 5 \\ 
    \cline{2-4}
    & Varied life & You can create music you wouldn’t have thought of, feeling a sense of novelty, and notice patterns in phrases you hadn’t realized. & 4 \\ 
    \cline{2-4}
    & Daring & You can pursue and challenge new genres and phrase patterns that you’ve never explored before. & 6 \\
    \hline

    Cybercode-1 & Pleasure & You feel happy because even things that are usually invisible or unrealistic can be made visible in a virtual space as the scenes you desire or want to display. & 0 \\ 
    \cline{2-4}
    & Enjoying life & You can enjoy fantasy by mixing elements that differ from reality, like in games, into your everyday environment. & 4 \\ 
    \cline{2-4}
    & Exciting life & For example, without visiting a museum, you can experience knowledge you previously knew only as text (e.g., from an encyclopedia) in 3D, allowing you to grasp the actual size of objects and have a realistic experience. & 3 \\ 
    \cline{2-4}
    & Varied life & It’s stimulating because you can experience a sense of escapism from everyday life. & 2 \\ 
    \cline{2-4}
    & Daring & You can explore dangerous places that are normally inaccessible in real life. & 7 \\ 
    \cline{2-4}
    & Choosing own goals & You can realize your desired space in reality, such as with interior design layouts. & 7 \\ 
    \cline{2-4}
    & Creativity & If other companies are not using this tool and only your company is, you can create unique exhibits, allowing many people to view objects that are too delicate or heavy to display in person. & 5 \\ 
    \cline{2-4}
    & Curious & It makes it easier to show those unfamiliar with smartphones which buttons to press, facilitating the use of new tools and broadening their interests. & 4 \\ 
    \cline{2-4}
    & Freedom & It enables those with mobility challenges to experience places and objects through video without physically traveling. & 6 \\
    \hline

    Bubble Click-1 & Influential & The location where the soap bubbles pop can be used like a fortune-telling game or omikuji (Japanese fortune slips). & 7 \\ 
    \cline{2-4}
    & Successful & Organized a soap bubble competition with friends, where the person whose bubble lasted the longest or flew the farthest received a prize, giving a sense of accomplishment. & 5 \\ 
    \cline{2-4}
    & Pleasure & By counting the number of popped bubbles, children can know how many bubbles they created. & 5 \\ 
    \cline{2-4}
    & Enjoying life & Used the function to track the position of bubbles and created a real-life shooting game with a toy gun to play with children. & 0 \\ 
    \cline{2-4}
    & Exciting life & At a club in Ibiza, the music and lighting changed in sync with the foam when bubbles were released. & 3 \\
    
\end{longtable}
\endgroup

\begingroup
\footnotesize 
\centering
\rowcolors{2}{white}{gray!15}
\begin{longtable}{>{\raggedright\arraybackslash}p{0.15\textwidth} >{\raggedright\arraybackslash}p{0.13\textwidth} >{\raggedright\arraybackslash}p{0.5\linewidth} c}
    \caption{Result of scene description and scoring by Technology Deployment.} \label{tab:scenes_deployment}\\
    \hline
    \rowcolor{white} 
    \textbf{Function} & \textbf{Single Values} & \textbf{Scenes} & \textbf{Score}\\
    \hline
    \endfirsthead
    
    \caption[]{-- Continued} \\
    \hline
    \rowcolor{white} 
    \textbf{Function} & \textbf{Single Values} & \textbf{Scenes} & \textbf{Score}\\
    \hline
    \endhead
    \hline
    \endlastfoot

    Omoiiro-1 & Self Respect & By choosing from a photo album, you can discover your own preferences and favorite colors that reflect your personality. & 3 \\
    \cline{2-4}
    & Creativity & Creators from different fields collaborate to create a piece (e.g., a painter and a ceramic artist), and the original colors can be extracted from the painter's work. & 5 \\ 
    \cline{2-4}
    & Curious & It sparks curiosity to wonder, "What color is this?" and to identify which part of the original photo the extracted color came from. & 2 \\ 
    \cline{2-4}
    & Freedom & Increases the freedom of design. & 0 \\ 
    \cline{2-4}
    & World of beauty & When stating that the color was inspired by the Earth, presenting a photo of the Earth as supporting evidence allows one to understand the intricate beauty of nature's colors (such as ultramarine instead of blue, or yamabuki instead of orange). & 4 \\
    \cline{2-4}
    & Unity with nature & By extracting colors from nature images, the resulting design can harmonize with the original image, creating a sense of unity with nature. & 4 \\
    \cline{2-4}
    & Broad minded & Colors that go beyond implicit biases related to gender (e.g., warm or vivid tones for women's clothing) or habitual choices of designers can be extracted for original designs. & 6 \\
    \cline{2-4}
    & Equality & Even people who lack confidence in their color sense can automatically extract colors just by selecting their favorite image. Non-designers may create more convincing color schemes than professional designers. & 7 \\ 
    \cline{2-4}
    & Inner harmony & Helps to understand the environment in which you feel calm, such as lighting or interior colors. & 3 \\ 
    \cline{2-4}
    & True friendship & Allows sharing of colors that only friends and family would recognize. & 4 \\
    \cline{2-4}
    & Mature love & By using memorable photos, the colors based on those scenes or recognizable only by friends and family lead to a deeper attachment or familiarity with the work. & 7 \\ 
    \cline{2-4}
    & Helpful & When out of design ideas, you can infinitely generate designs. & 5 \\
    \cline{2-4}
    & Spiritual life & When there is an image supporting the design, a story emerges, making it more than just an object, but also providing spiritual support. & 6 \\
    \hline

    Omoiiro-2 & Pleasure & Even if you cannot design on your own, you can create a color palette from a photo and convey those colors to a designer to achieve the desired design. & 6 \\ 
    \cline{2-4}
    & Enjoying life & You can feel the changing of the seasons through a palette generated from photos. & 5 \\ 
    \cline{2-4}
    & Exciting life & The excitement comes from generating a color palette from an original image, creating colors you had not imagined. & 5 \\ 
    \cline{2-4}
    & Varied life & By using images of the four seasons to create a palette, you can decorate a store in accordance with the season. & 7 \\ 
    \cline{2-4}
    & Creativity & Students who are creating something but have not yet grasped their own unique style can take reference photos and, through comparison with other works, find their own distinct color identity. & 4 \\ 
    \cline{2-4}
    & Curious & When colors you've never seen before are displayed, it sparks your curiosity. & 5 \\ 
    \cline{2-4}
    & Freedom & Expanding creative horizons by being forced to use colors from a palette that you wouldn't normally choose. & 0 \\ 
    \cline{2-4}
    & Self Respect & Seeing various color variations in a palette increases your range of ideas, leading to more opportunities to express your true self. & 2 \\ 
    \cline{2-4}
    & Choosing own goals & Each variation in the color palette hints at different possibilities for your future. & 0 \\ 
    \cline{2-4}
    & Independent & You can take on the task of creating a palette. & 2 \\
    \hline

    Continuator-1 & Pleasure & Amateur or student musicians who are unsure of their core performance style will be happy to rediscover their essence through this feedback. & 5 \\ 
    \cline{2-4}
    & Enjoying life & By listening to the music you play casually every day—whether in the morning or after a shower—you can rediscover the core aspects of your performance. & 5 \\ 
    \cline{2-4}
    & Exciting life & Discovering performance traits you were previously unaware of, providing an exciting stimulus. & 7 \\ 
    \cline{2-4}
    & Varied life & By listening to music that differs from your original work, you can experience a refreshing sense of novelty. & 7 \\ 
    \cline{2-4}
    & Self Respect & Extracting traits that you lacked confidence in can lead to a sense of validation and self-affirmation. & 4 \\ 
    \cline{2-4}
    & Choosing own goals & By identifying clear weaknesses in your tendencies, you can set specific goals for improvement. & 4 \\ 
    \cline{2-4}
    & Creativity & By comparing your extracted musical traits with others, you can understand what makes your work unique. & 6 \\ 
    \cline{2-4}
    & Curious & Discovering traits you were previously unaware of sparks curiosity and interest. & 6 \\ 
    \cline{2-4}
    & Freedom & Feeling validated in areas where you lacked confidence gives you the freedom to play with more confidence in the future. & 0 \\ 
    \cline{2-4}
    & Independent & You can discover your core musical identity without relying on teachers or others. & 3 \\ 
    \hline

    Continuator-2 & Pleasure & When a composer feels limited in inspiration and dissatisfied with an incomplete composition, it can offer partial reworks and suggestions. Joy and satisfaction arise when better phrases and developments are found. & 3 \\ 
    \cline{2-4}
    & Enjoying life & The generated music maintains a consistent atmosphere, making it ideal for background music. & 2 \\ 
    \cline{2-4}
    & Exciting life & Like the solo part of jazz, real-time improvisation enables collaboration between humans and machines. & 6 \\ 
    \cline{2-4}
    & Varied life & Using a famous 3-minute song for a dance. While the first 3 minutes are familiar, the music continues with new, unheard phrases, allowing real-time choreography that gives the audience a sense of novelty. & 4 \\ 
    \cline{2-4}
    & Daring & Stimulating phrases emerge that you wouldn’t normally think of, inspiring entirely new melody lines. & 6 \\ 
    \cline{2-4}
    & Self Respect & Your personal style is preserved in the continuation of your own music. & 0 \\ 
    \cline{2-4}
    & Creativity & If you're daily vlogging, it can generate music in real-time that matches the content, turning ordinary moments into more unique content for viewers. & 6 \\ 
    \cline{2-4}
    & Curious & After listening to several performances generated by the Continuator, you'll want to try playing in different patterns to see what new melodies emerge. & 4 \\ 
    \cline{2-4}
    & Independent & When making a wedding video, the selected BGM was slightly shorter than the footage. By automatically generating a continuation, the music could be extended to the perfect length without relying on others. & 7 \\
    \hline

    Cybercode-1 & Pleasure & You can summon your beloved subjects (idols, pets, game content, etc.) in real time. & 5 \\ 
    \cline{2-4}
    & Enjoying life & You can save cherished memorabilia that are too bulky to keep, along with their original locations. & 7 \\ 
    \cline{2-4}
    & Exciting life & Everyday places like your commuting or school route become key spots in a game through AR, creating an exciting experience. & 5 \\ 
    \cline{2-4}
    & Varied life & By adding more codes, you can introduce additional CG characters. & 3 \\ 
    \cline{2-4}
    & Daring & Searching for hidden items and visiting places you wouldn't normally go to retrieve something only available there. & 5 \\ 
    \cline{2-4}
    & Curious & Curiosity arises from wondering what will be generated from this code. It serves as a marker for what ARCG will appear. & 3 \\ 
    \cline{2-4}
    & Freedom & You can place art in public spaces without causing any disturbance. & 0 \\
    \hline

    Bubble Click-1 & Social recognition & By being able to predict where soap bubbles will pop, you can gain praise from friends around you. & 5 \\ 
    \cline{2-4}
    & Influential & Expanding the technology to applications like concrete or medical devices can improve the performance of bubble sensors. & 6 \\ 
    \cline{2-4}
    & Successful & By quantifying and visualizing things that were previously invisible, research can progress, leading to new discoveries. & 7 \\ 
    \cline{2-4}
    & Capable & Having measurement data allows bubble companies to use it for developing longer-lasting soap bubbles. & 5 \\ 
    \cline{2-4}
    & Enjoying life & You can create a new form of entertainment by challenging how long a soap bubble can fly. & 0 \\
    
\end{longtable}
\endgroup

\section*{Appendix C. Detailed Results of workshops: technology-side evaluation}\label{app:ws_trl_results}

This appendix presents the detailed results of the technology-side evaluation, as described in Section~\ref{subsec:trl_assessment}.
The following tables show the Technology Readiness Level (TRL) scores assigned by the expert from Sony CSL.
The evaluation was conducted only for the use-case scenes that received a market-side score of $5$ or higher.

Table \ref{tab:trl_consumers} shows the results for the scenes generated by the General Consumers group.
Table \ref{tab:trl_deployment} shows the results for the scenes generated by the Technology Deployment group.

\begingroup
\footnotesize 
\centering
\rowcolors{2}{white}{gray!15}
\begin{longtable}{>{\raggedright\arraybackslash}p{0.15\textwidth} >{\raggedright\arraybackslash}p{0.13\textwidth} >{\raggedright\arraybackslash}p{0.5\linewidth} c}
    \caption{Result of technology-side evaluation by general consumer.}\label{tab:trl_consumers}\\
    \hline
    \rowcolor{white} 
    \textbf{Function} & \textbf{Single Values} & \textbf{Scenes} & \textbf{TRL}\\
    \hline
    \endfirsthead
    
    \caption[]{-- Continued} \\
    \hline
    \rowcolor{white} 
    \textbf{Function} & \textbf{Single Values} & \textbf{Scenes} & \textbf{TRL}\\
    \hline
    \endhead
    \hline
    \endlastfoot

    Omoiiro-1 & Curious & I created a puzzle where you use the colors of a palette to make many small monochrome square pieces, and then recreate the original image using pointillism. & 6 \\
    \cline{2-4}
    & Broad minded & Together with some friends, we each guessed the color we thought best represented the atmosphere of an image, discussed the reasons, and then compared our choices with the palette extracted by the program for feedback. & 6 \\
    \cline{2-4}
    & Equality & By visualizing issues like educational inequality or gender disparities through color, it deepens understanding of the problems and inspires action toward solutions. & 1 \\
    \cline{2-4}
    & World of beauty & I extracted colors from a photo of nature. The photo contained more colors than I expected, and seeing previously unrecognized colors revealed hidden beauty in nature. & 6 \\
    \hline

    Omoiiro-2 & Enjoying life & By converting a photo you took into a palette, you can express your current feelings, share it on social media, and feel a sense of connection if someone else has a similar color palette. & 6 \\
    \cline{2-4}
    & Exciting life & Input your mood for the day, and based on the recommended palette, you can choose clothes that reflect that mood, leading to an exciting day as you go out wearing them. & 1 \\
    \cline{2-4}
    & Varied life & You can more clearly feel the changing of the seasons through the shift in color palettes, and reflect this in your hair color or clothing coordination. & 1 \\
    \hline

    Continuator-1 & Successful & A person who wants to compose music can create their desired song and feel a sense of accomplishment. & 6 \\
    \cline{2-4}
    & Capable & A person who wants to compose music can efficiently move closer to their desired song by extracting characteristics from hit songs, among other things. & 6 \\
    \cline{2-4}
    & Pleasure & Even if someone wants to compose music but cannot do it well, they can compose just by inputting their desired song image at the beginning. They can feel joy as their wish to compose music is fulfilled. & 6 \\
    \cline{2-4}
    & Exciting Life & General consumers can have an exciting experience by creating their own music. & 6 \\
    \cline{2-4}
    & Varied Life & A person who wants to compose music can create different types of songs without repeating similar phrases. They can create songs with different features and new images. & 6 \\
    \hline

    Continuator-2 & Pleasure & You feel joy in composing music, even without any skills or prior experience. & 6 \\
    \cline{2-4}
    & Enjoying life & You find enjoyment in the ease of composing music as part of your everyday life. & 6 \\
    \cline{2-4}
    & Exciting life & Unexpected songs may emerge or be performed, making it a stimulating experience. & 6 \\
    \cline{2-4}
    & Daring & You can pursue and challenge new genres and phrase patterns that you’ve never explored before. & 6 \\
    \hline

    Cybercode-1 & Daring & You can explore dangerous places that are normally inaccessible in real life. & 9 \\
    \cline{2-4}
    & Choosing own goals & You can realize your desired space in reality, such as with interior design layouts. & 9 \\
    \cline{2-4}
    & Creativity & If other companies are not using this tool and only your company is, you can create unique exhibits, allowing many people to view objects that are too delicate or heavy to display in person. & 9 \\
    \cline{2-4}
    & Freedom & It enables those with mobility challenges to experience places and objects through video without physically traveling. & 9 \\
    \hline

    Bubble Click-1 & Influential & The location where the soap bubbles pop can be used like a fortune-telling game or omikuji (Japanese fortune slips). & 3 \\
    \cline{2-4}
    & Successful & Organized a soap bubble competition with friends, where the person whose bubble lasted the longest or flew the farthest received a prize, giving a sense of accomplishment. & 3 \\
    \cline{2-4}
    & Pleasure & By counting the number of popped bubbles, children can know how many bubbles they created. & 4 \\
    
\end{longtable}
\endgroup

\begingroup
\footnotesize 
\centering
\rowcolors{2}{white}{gray!15}
\begin{longtable}{>{\raggedright\arraybackslash}p{0.15\textwidth} >{\raggedright\arraybackslash}p{0.13\textwidth} >{\raggedright\arraybackslash}p{0.5\linewidth} c}
    \caption{Result of technology-side evaluation by Technology Deployment.} \label{tab:trl_deployment}\\
    \hline
    \rowcolor{white} 
    \textbf{Function} & \textbf{Single Values} & \textbf{Scenes} & \textbf{TRL}\\
    \hline
    \endfirsthead
    
    \caption[]{-- Continued} \\
    \hline
    \rowcolor{white} 
    \textbf{Function} & \textbf{Single Values} & \textbf{Scenes} & \textbf{TRL}\\
    \hline
    \endhead
    \hline
    \endlastfoot

    Omoiiro-1 & Creativity & Creators from different fields collaborate to create a piece (e.g., a painter and a ceramic artist), and the original colors can be extracted from the painter's work. & 9 \\ 
    \cline{2-4}
    & Broad minded & Colors that go beyond implicit biases related to gender (e.g., warm or vivid tones for women's clothing) or habitual choices of designers can be extracted for original designs. & 9 \\
    \cline{2-4}
    & Equality & Even people who lack confidence in their color sense can automatically extract colors just by selecting their favorite image. Non-designers may create more convincing color schemes than professional designers. & 6 \\ 
    \cline{2-4}
    & Mature love & By using memorable photos, the colors based on those scenes or recognizable only by friends and family lead to a deeper attachment or familiarity with the work. & 6 \\ 
    \cline{2-4}
    & Helpful & When out of design ideas, you can infinitely generate designs. & 9 \\
    \cline{2-4}
    & Spiritual life & When there is an image supporting the design, a story emerges, making it more than just an object, but also providing spiritual support. & 6 \\
    \hline

    Omoiiro-2 & Pleasure & Even if you cannot design on your own, you can create a color palette from a photo and convey those colors to a designer to achieve the desired design. & 9 \\
    \cline{2-4}
    & Enjoying life & You can feel the changing of the seasons through a palette generated from photos. & 9 \\
    \cline{2-4}
    & Exciting life & The excitement comes from generating a color palette from an original image, creating colors you had not imagined. & 9 \\
    \cline{2-4}
    & Varied life & By using images of the four seasons to create a palette, you can decorate a store in accordance with the season. & 9 \\
    \cline{2-4}
    & Curious & When colors you've never seen before are displayed, it sparks your curiosity. & 9 \\
    \hline
    
    Continuator-1 & Pleasure & Amateur or student musicians who are unsure of their core performance style will be happy to rediscover their essence through this feedback. & 6 \\
    \cline{2-4}
    & Enjoying life & By listening to the music you play casually every day—whether in the morning or after a shower—you can rediscover the core aspects of your performance. & 6 \\
    \cline{2-4}
    & Exciting life & Discovering performance traits you were previously unaware of, providing an exciting stimulus. & 6 \\
    \cline{2-4}
    & Varied life & By listening to music that differs from your original work, you can experience a refreshing sense of novelty. & 6 \\
    \cline{2-4}
    & Creativity & By comparing your extracted musical traits with others, you can understand what makes your work unique. & 3 \\
    \cline{2-4}
    & Curious & Discovering traits you were previously unaware of sparks curiosity and interest. & 6 \\ 
    \hline

    Continuator-2 & Exciting life & Like the solo part of jazz, real-time improvisation enables collaboration between humans and machines. & 6 \\
    \cline{2-4}
    & Daring & Stimulating phrases emerge that you wouldn’t normally think of, inspiring entirely new melody lines. & 6 \\
    \cline{2-4}
    & Creativity & If you're daily vlogging, it can generate music in real-time that matches the content, turning ordinary moments into more unique content for viewers. & 4 \\
    \cline{2-4}
    & Independent & When making a wedding video, the selected BGM was slightly shorter than the footage. By automatically generating a continuation, the music could be extended to the perfect length without relying on others. & 4 \\
    \hline

    Cybercode-1 & Pleasure & You can summon your beloved subjects (idols, pets, game content, etc.) in real time. & 9 \\
    \cline{2-4}
    & Enjoying life & You can save cherished memorabilia that are too bulky to keep, along with their original locations. & 8 \\
    \cline{2-4}
    & Exciting life & Everyday places like your commuting or school route become key spots in a game through AR, creating an exciting experience. & 9 \\
    \cline{2-4}
    & Daring & Searching for hidden items and visiting places you wouldn't normally go to retrieve something only available there. & 9 \\
    \hline

    Bubble Click-1 & Social recognition & By being able to predict where soap bubbles will pop, you can gain praise from friends around you. & 5 \\
    \cline{2-4}
    & Influential & Expanding the technology to applications like concrete or medical devices can improve the performance of bubble sensors. & 2 \\
    \cline{2-4}
    & Successful & By quantifying and visualizing things that were previously invisible, research can progress, leading to new discoveries. & 4 \\
    \cline{2-4}
    & Capable & Having measurement data allows bubble companies to use it for developing longer-lasting soap bubbles. & 1 \\

\end{longtable}
\endgroup



\end{document}